\documentclass[floatfix,aps,prb,twocolumn,groupedaddress]{revtex4}

\usepackage{epsfig}

\usepackage{color}

\newcommand{\be}{\begin{equation}}
\newcommand{\ee}{\end{equation}}
\newcommand{\bea}{\begin{eqnarray}}
\newcommand{\eea}{\end{eqnarray}}

\begin{document}


\title{Lattice defects and boundaries in conducting carbon nanotubes}


\author{Sebastian A. Reyes$^{1,2}$, Alexander Struck$^1$ and Sebastian Eggert$^1$}
\affiliation{$^1$Department of Physics and Research Center OPTIMAS, University of Kaiserslautern, D-67663, Kaiserslautern, Germany}
\affiliation{$^2$Facultad de F\'isica, Pontificia Universidad Cat\'olica de Chile, Casilla 306, Santiago 22, Chile}

\pacs{PACS numbers:  73.22.-f, 61.48.-c, 71.20.Tx, 73.61.Wp}

\date{\today}

\begin{abstract}

We consider the effect of various defects and boundary structures on the
low energy electronic properties in conducting zigzag and armchair carbon nanotubes.
The tight binding model of the conduction bands is mapped exactly
onto
simple lattice models consisting of two uncoupled parallel chains.
Imperfections such as impurities, structural defects or caps can be
easily included into the effective lattice models, allowing  a detailed physical
interpretation of their consequences.
The method is quite general and can be used to study a wide range of possible
imperfections in carbon nanotubes. We obtain the electron density
patterns expected from a scanning tunneling microscopy experiment for half
fullerene caps and typical impurities in the bulk of a tube,
namely the Stone-Wales defect and a single vacancy.
\end{abstract}

\pacs{}

\maketitle


\section{Introduction}

The quasi one dimensional nature and interesting electronic
properties of single-walled carbon nanotubes (SWCNTs)
have attracted a lot of attention ever since they where first synthesized
in 1993.\cite{IijimaSW, Bethune} The tubes are well described by a
graphene sheet that is rolled up along a wrapping vector
(${\bf C}_h = N{\bf a}_1 + M{\bf a}_2$) with characteristic chiral indices
$(N, M)$.\cite{book}  The finite circumference results in discrete one-dimensional bands, which may
be conducting only if they intersect the Fermi points.
In case that $(N-M)$ is divisible by three, there are exactly two such conduction bands
while otherwise the tube will be
semiconducting.
Strain and the intrinsic curvature slightly shift the Fermi points, which may open a
very small ''semi-metallic'' gap,
but otherwise leave the electronic structure intact.\cite{Kane,gap}
Due to their large aspect ratio and very high structural
quality, SWCNTs are often modeled as having infinite length and being completely
free of defects. In many situations though, such approximations are not entirely justified.

For example, tubes can be cut into lengths well below 100 nm allowing for clear
experimental observation of phenomena related to the one dimensional confinement
of electrons.\cite{Rubio, Rochefort, Jiang, Venema, Lemay}
Any slight deviations from a simple
``particle-in-a-box'' picture may be due to interaction effects\cite{anfuso,schneider}
or else must be due to non-trivial boundary conditions at the ends.
Semi-infinite tubes have also been shown to have an interesting oscillating
electronic structure near the ends which even carry some signatures
of interactions.\cite{lee,stm}
However, the tube ends are usually not perfect and may contain a number of
defects or can even be closed by structures such as a half fullerene,
the effect of which we will consider
explicitly in this paper. Moreover, it is
known that even high-quality SWCNTs contain on average one defect every
4 $\mu$m.\cite{Fan} Imperfections in the bulk such as
lattice deformations, vacancies or ad-atoms may be induced by strain or
irradiation and can drastically modify their electronic and transport
properties.\cite{Charlier, Chico, Choi, Bockrath, Kong_transp, Biel_And_loc, Gomez_And_loc}
Such irregularities can then potentially be used to tailor the properties of CNTs
\cite{Son, Rocha, Park} for its use in a wide range of applications in
nanodevices.\cite{Bachtold, Carb_elec} In this context, precise understanding
of the consequences of impurities and structural defects becomes especially
important. Moreover, the study of disorder \cite{Mattis} and single
impurities \cite{KaneFisher,EggertAffleck} in one-dimensional systems remains
of fundamental interest.  It has been shown that
impurities in general give rise to characteristic oscillations of the tunneling
density of states
that reflect the symmetry breaking around the defects,\cite{kane3} albeit without considering
explicit microscopic models of the impurities on the lattice.  The amplitudes of
density oscillations are in turn
related to the scattering strength and the resulting conductivity.\cite{rommer}

In the present work we derive effective lattice models of uncoupled chains 
within the tight binding picture of the lowest conduction band.
The simple models allow the systematic 
study of the effect from various 
defects in the bulk as well as edge structures
on the electronic structure of otherwise perfect tubes.
In particular, we
focus on impurity effects in the conduction bands of finite
metallic SWCNTs with open ends.
The non-conducting bands of metallic SWCNTs are energetically separated by
a semiconducting gap that is inversely proportional to the radius\cite{Kane}
($\Delta \sim 1/|\vec{C_h}|$) on the order of 0.5 eV for a radius of $7\rm \AA$.  
In this work only 
excitations below this energy scale will be considered which are believed to be well 
described by the conduction bands only.  The tight binding structure
of the conduction bands in zig-zag and armchair carbon nanotubes can be
captured exactly by simplified lattice models for SWCNTs.\cite{BalFish,Lin}
 Our derivation results in a lattice of two decoupled chains, which allows
the incorporation of simple defects or even more complicated
structures such as caps by adding
a small number of extra sites. It is demonstrated that, depending on the nature of
the imperfection under consideration, the initially independent chains can either
remain independent or become connected by the presence of the defect.
Patterns for the local density of states (LDOS) near the Fermi level are
obtained using such models and a straight-forward physical interpretation
is possible due to the simplicity of the effective theory. 
At this point
it is worth mentioning that such a method accurately reproduces the density
patterns for the perfect tube obtained by using first-principles calculations
in Ref.~[\onlinecite{Rubio}] and the results are also consistent with the
scanning tunneling microscopy (STM) experiments reported in Ref.~[\onlinecite{Venema}].
As such, the simple lattice models here capture the most relevant physical 
effects of defects, but for quantitative details future comparison with 
first principle calculations would also be useful.  In particular, 
while our exemplary vacancy model only illustrates the first order symmetry
breaking effect, it would also 
be interesting to incorporate the relaxation of the structure around
 defects.\cite{relax} While this can in principle be considered 
within the same formalism,  this extension would require additional 
input from independent ab-initio calculations about the modified hopping parameters.

Electron-electron interaction effects can also be incorporated into the
simple lattice models as described in the appendix \ref{Interactions}, but are
ignored for the calculation of the STM patterns from defect scattering to
first order.  This is justified because in STM experiments
the Coulomb interaction becomes short range due to screening from the
substrate and the effective
on-site interaction $U_{\rm eff}=U/N$ is weak for large enough $N$.\cite{BalFish}
It is known that interaction effects result in a characteristic amplitude modulation
of the scattered density waves,\cite{anfuso,schneider,stm,lee} but appear to leave
the principle wave pattern intact.
Nevertheless, it should be emphasized that the models presented here, are well
suited for the study of strong correlations by applying the powerful techniques
available for one-dimensional systems such as bosonization \cite{GogolinNerTsv} or
density matrix renormalization group (DMRG).\cite{White} For the interested reader,
all additional terms necessary to take electron-electron interactions into account
in the effective models are listed in Appendix \ref{Interactions}.

The paper is organized as follows: In Sec.~\ref{Eff_Mod} we show how the effective theory for the perfect CNTs is obtained. Then, in Sec.~\ref{boundaries} we incorporate 
imperfections and caps at the tube ends. We proceed similarly in Sec.~\ref{surf_def} but 
now considering structural deformations in the bulk of the tube. For each case, we use the effective models to obtain electron densities in real space for states near the Fermi level. The resulting patterns are the ones expected for a typical STM experiment.\cite{Venema,Lemay}

\section{Effective Models} \label{Eff_Mod}

The tight binding band structure of an infinite graphene sheet includes a valence band and a
conduction band that touch at two distinct Fermi points.
When this graphite monolayer is rolled into a CNT,
boundary conditions are imposed such that only a finite set of bands are allowed. As 
mentioned above, in some cases a subset of bands coincides with the Fermi points and the tube is metallic. The present section is devoted to show how to obtain an effective low energy theory for conducting SWCNTs by keeping only its conduction bands. The result is a dramatically simplified description of the tubes in terms of two chain models. To be able to take advantage of these simple lattice models, we do {\it not} need to recur to the
widely used linearization around the Fermi points.\cite{Kane,Egger,Kane2,kane3} The procedure will be outlined for the armchair and conducting zig-zag nanotubes for which the conduction bands can be easily identified and isolated once a Fourier transform around the tube is performed.

\subsection{Armchair CNT}

\begin{figure}
\centering
\includegraphics[width=0.44\textwidth]{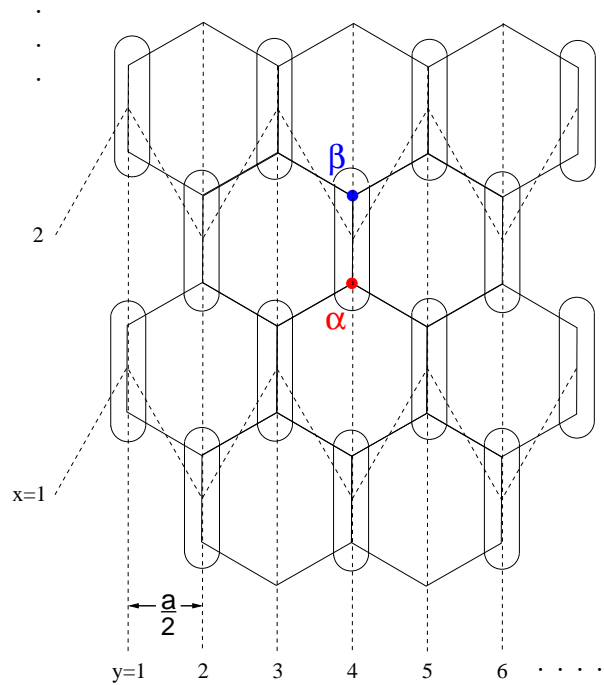}
\caption{(Color online) ARMCHAIR: Labels chosen for the atoms in the graphene lattice in the case of the armchair CNT. The hexagonal network is composed by two triangular sublattices which are labeled $\alpha$ and $\beta$. Within the tight binding approximation there is a hopping amplitude $t$ between them. The label $y$ is in the direction along the tube axis, 
and $x$ indicates the bonds around the tube.}\label{ACH_CNT}
\end{figure}

Armchair carbon nanotubes are always conducting and are defined by a chiral vector with equal indices $(N, N)$.\cite{book}
As usual, the atoms can be divided into two distinct sets $\alpha$ and $\beta$, corresponding to the two sublattices of the graphene honeycomb structure. It is convenient to label each of the $\alpha$-$\beta$ pairs 
with a set of coordinates $(x, y)$ as shown in Fig.~\ref{ACH_CNT}.
The coordinate $y$ is along the tube and $x$ is around the tube. 
The tight binding Hamiltonian can now be written as follows
\bea
H  =  -t \sum_{x=1}^{N} &  & \left( \sum_{y ~odd}^{~}  \left(
\alpha^{\dagger}_{x, y}\beta^{\phantom\dagger}_{x, y+1} +
\beta^{\dagger}_{x, y}\alpha^{\phantom\dagger}_{x+1, y+1} \right)
\right. \nonumber \\     + &&\left. 
\sum_{y ~even}^{~}  \left(\beta^{\dagger}_{x, y}\alpha^{\phantom\dagger}_{x, y+1} + 
\alpha^{\dagger}_{x, y}\beta^{\phantom\dagger}_{x-1, y+1} \right) 
\right. \nonumber \\     +& &\left. 
\sum_{y=1}^{L} \alpha^{\dagger}_{x, y}\beta^{\phantom\dagger}_{x, y} 
\right) + h.c.   \label{H_arm}
\eea
Here $\alpha$ and $\beta$ are the destruction operators in the corresponding sublattice, $L$ is the length of the tube, and $N$ is the number of bonds around its perimeter.
Performing a partial Fourier transform in the direction $x$ around the tube and considering that the only conducting modes are the
ones corresponding to $k=0$ (one for each sublattice),\cite{book} the following approximation
holds
\bea
\alpha_{x,y} = \frac{1}{\sqrt{N}} \sum_{k} \alpha_{k,y} e^{ikx} \approx \frac{1}{\sqrt{N}} \alpha_{k=0,y}, ~~~ k = \frac{2\pi l}{N},  \label{FT_armchair}
\eea
and analogously for  $\beta$. Since the $k=0$ mode is the only one that will be taken into account, we drop the $k$ index in the following. Performing the linear transformation
\bea
\alpha_{y} = \frac{S_{y} + A_{y}}{\sqrt{2}} , ~~~~~ \beta_{y} = \frac{S_{y} - A_{y}}{\sqrt{2}},    \label{LT_armchair}
\eea
the effective conduction band Hamiltonian can be written in terms of two completely independent parts, each one of them involving only the symmetric (S) or anti-symmetric (A) mode
\bea
H_{\rm eff} = H_S + H_A . \label{eff_H_arm}
\eea
\begin{figure}
\centering
(a) \includegraphics[width=0.44\textwidth]{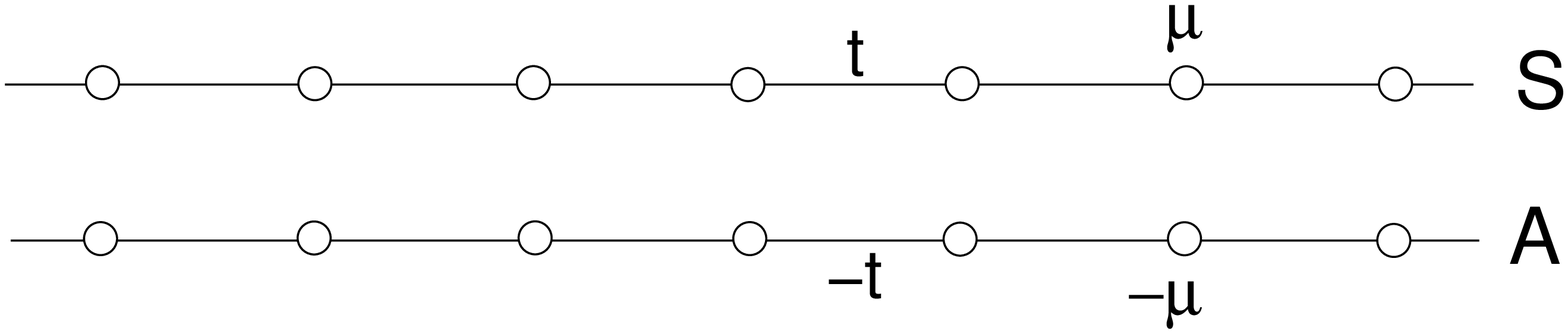}  \\
(b) \includegraphics[width=0.44\textwidth]{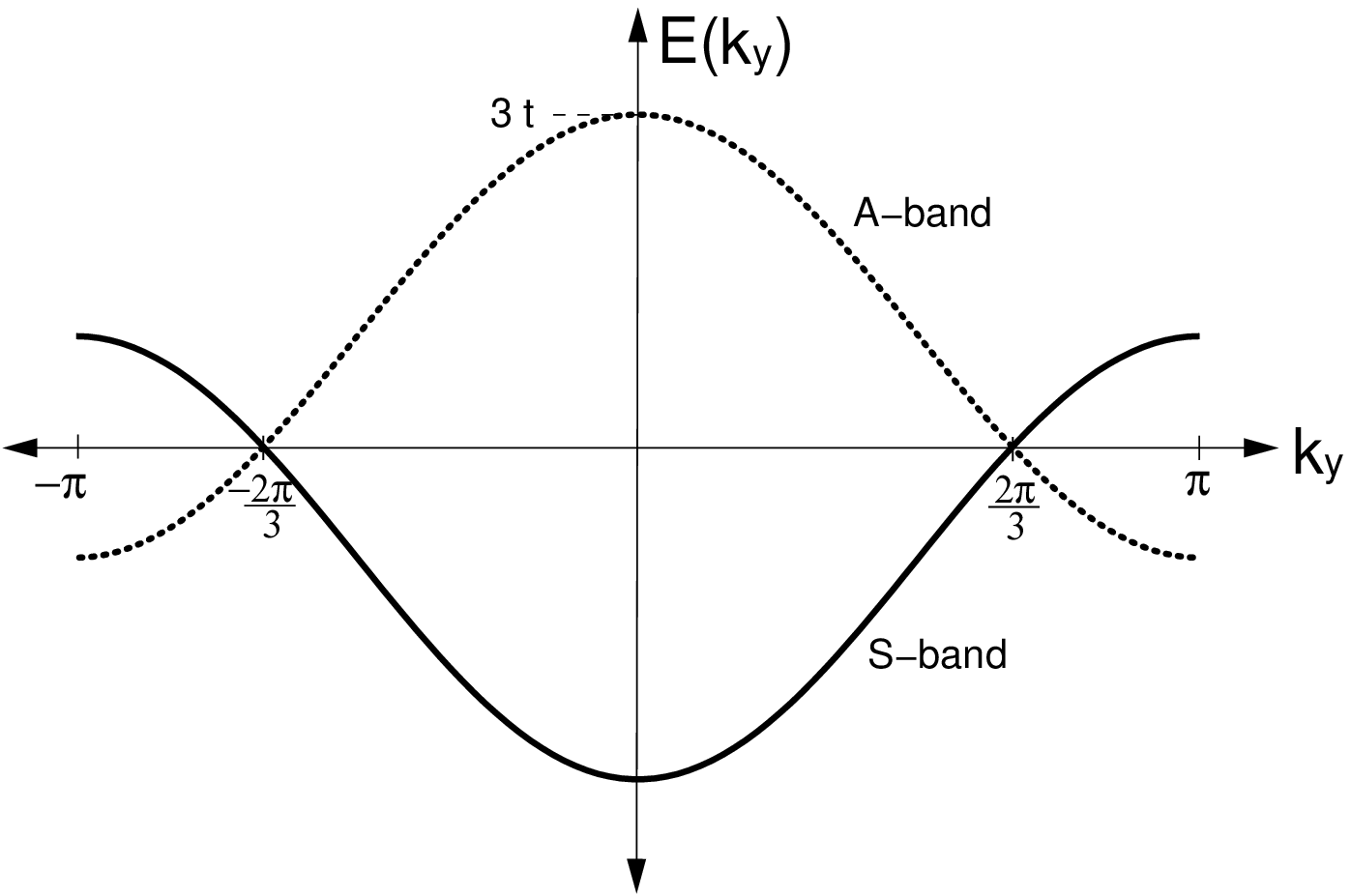}
\caption{(a) Effective chain lattice model for the perfect armchair carbon nanotube. The upper chain corresponds to the symmetric ($S$) mode and the lower one to the anti-symmetric ($A$) mode. In this case $\mu=t$. (b) Band structure for the conducting modes
(effective lattice spacing $a_{\rm eff}=a/2$). }\label{double_chain2}
\end{figure}
Inserting Eqs.~(\ref{FT_armchair}) and (\ref{LT_armchair}) into Eq.~(\ref{H_arm}) we find
\bea
H_S &=& -t \sum_{y}  \left( S_{y}^{\dagger}S^{\phantom\dagger}_{y+1} + \frac{1}{2}S_{y}^{\dagger}S^{\phantom\dagger}_{y} + h.c. \right)   
\\ \nonumber &=& 
 -t \sum_{k_y} \left(2 \cos{k_y} + 1 \right) S_{k_y}^{\dagger}S^{\phantom\dagger}_{k_y}  \\
H_A &=& t \sum_{y} \left( A_{y}^{\dagger}A^{\phantom\dagger}_{y+1} + \frac{1}{2}A_{y}^{\dagger}A^{\phantom\dagger}_{y} + h.c. \right) 
\\ \nonumber &=& 
t \sum_{k_y} \left(2 \cos{k_y} + 1 \right)  A_{k_y}^{\dagger}A^{\phantom\dagger}_{k_y}
\eea
Therefore, by keeping only the degrees of freedom of the conduction bands,
the effective lattice model consists of two decoupled chains as depicted in 
Fig.~\ref{double_chain2}.   

Possible more realistic longer range hopping
could also be included using this formalism.    
In particular, it is also possible to use the exact Wannier orbitals
for the basis $\alpha_{x,y}$ and $\beta_{x,y}$, e.g.~if they are determined 
from ab-initio methods.  This in turn would give additional 
longer range hopping terms in the Hamiltonian (\ref{H_arm}) of the conduction band.
Since the symmetry is not changed, the transformations
in Eqs.~(\ref{FT_armchair}) and (\ref{LT_armchair}) are still equally valid and yield 
again two decoupled chains albeit with corresponding 
longer range hopping within each chain.  For the hybridization due to curvature 
approximate $\alpha_{x,y}$ and $\beta_{x,y}$ orbitals and the corresponding 
change in hopping
have been predicted for the tight binding model\cite{kleiner}
which are known not to change the symmetry or the electronic structure 
for armchair tubes.\cite{Kane,gap,kleiner}

\subsection{Zigzag CNT}
\begin{figure}
\centering
\includegraphics[width=0.44\textwidth]{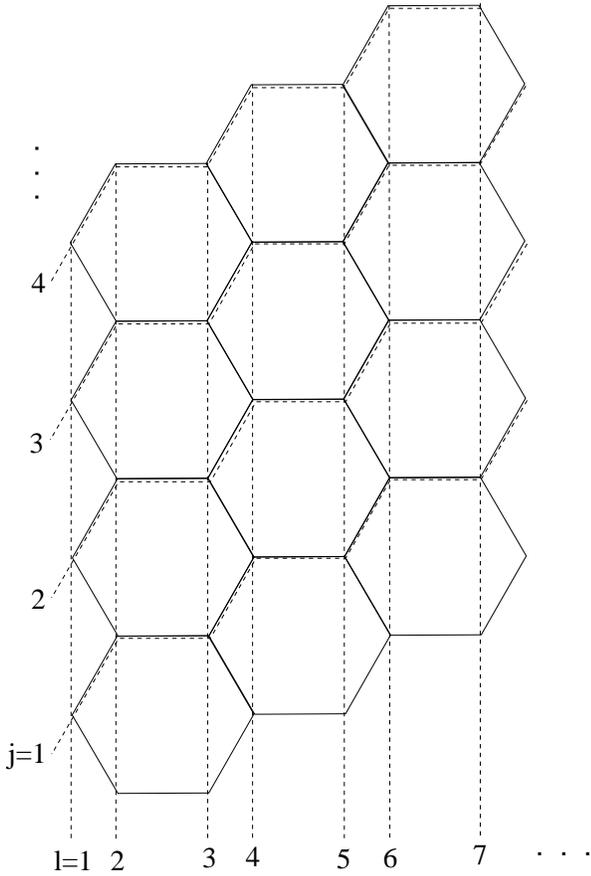}
\caption{(Color online) ZIGZAG: Labeling scheme used for the zigzag CNT. Coordinate $j$ marks the sites around the tube. The two different sublattices $\alpha$ and $\beta$,
now correspond to even and odd values of the coordinate $l$ along the tube axis,
respectively.}\label{ZIG-ZAG}
\end{figure}

For the zigzag tube we employ the labeling of the atoms shown in Fig.~\ref{ZIG-ZAG}.
The chiral indices of the wrapping vector of the zigzag tubes is given by
$(N,0)$ where $Na$ is the circumference of the tube.
Using the notation from Fig.~\ref{ZIG-ZAG} the Hamiltonian can be written
\bea
H &=& -t \sum_{l}^{} \sum_{j}^{}\left( c^{\dagger}_{j,l}c^{\phantom\dagger}_{j,l+1} + h.c.\right) 
\\ \nonumber && 
-t \sum_{l~{\rm odd}}^{} \sum_{j}^{}\left( c^{\dagger}_{j,l}c^{\phantom\dagger}_{j-1,l+1} + h.c. \right)
\eea
Introducing the Fourier transform of the fermionic operators in the direction around the tube
\bea
c_{j,l}^{\dagger} = \frac{1}{\sqrt{N}} \sum_{k} c_{k,l}^{\dagger} e^{-{\rm i} kj} , ~~~k=\frac{2\pi n}{N}    \label{FT_zigzag}
\eea
the Hamiltonian becomes
\bea
H = -t \sum_{k} & & \!\!\!\!\!\!\left\{ \sum_{l}^{}\left( c^{\dagger}_{k,l}c^{\phantom\dagger}_{k,l+1} + h.c. \right)
\right. \\ \nonumber   &&\left. 
+ \sum_{l~{\rm odd}}^{} \left(e^{-{\rm i}k}c^{\dagger}_{k,l}c^{\phantom\dagger}_{k,l+1} + h.c. \right) \right\}
\eea
Only zigzag tubes with $N$ divisible by three are metallic and the
conduction bands are given by the two wavevectors $k=\pm Q=\pm\frac{2\pi}{3}$
around the tube.\cite{book}  
Restricting the Fourier sum accordingly and using
$1 + e^{\pm {\rm i} \frac{2\pi}{3}} = e^{\pm {\rm i} \frac{\pi}{3}},$
the Hamiltonian for the conduction bands becomes
\bea
H = -t \sum_{k=\pm Q} & & \!\!\!\!\!\! \left\{ \sum_{l~ \rm even}^{}\left( 
c^{\dagger}_{k,l}c^{\phantom\dagger}_{k,l+1} + h.c. \right) 
\right.\\ \nonumber &&\left. 
 + \sum_{l~\rm odd}^{}\left( e^{-{\rm i}\frac{k}{2}}c^{\dagger}_{k,l}c^{\phantom\dagger}_{k,l+1} + h.c. \right)\right\} 
\eea
Further simplification is achieved if one introduces the transformation
\bea
c_{\pm Q,l} = e^{\pm {\rm i} \phi(l)}C_{\pm,l}  \label{C_transf}
\eea
where
\bea
\phi(l) = \frac{\pi}{6} \left[l+(l+1)(\mathrm{mod}~ 2)\right]  \label{phi_l}
\eea
which results in a simple two band nearest neighbor hopping model
\bea
H_{{\rm eff}} = -t \sum_{\mu=\pm}  \sum_{l}^{}\left( C^{\dagger}_{\mu,l}C^{\phantom\dagger}_{\mu,l+1} + h.c. \right) \label{ZZ_eff_H} 
\eea
By restricting the degrees of freedom to the conduction bands,
the creation operators on the lattice of the nano\-tube are therefore reduced to
\bea
c_{j,l}^{\dagger} = \frac{1}{\sqrt{N}} \left( C_{+,l}^{\dagger} e^{- {\rm i} (\phi(l)+Qj)} + C_{-,l}^{\dagger} e^{ {\rm i} (\phi(l)+Qj)} \right). \label{C_reverse}
\eea
As mentioned above, even and odd chain sites $l$ now correspond to the two different
sublattices of the tube, while the two chains correspond to the two allowed
wavevectors
$k=\pm Q=\pm\frac{2\pi}{3}$ around the tube which are exactly degenerate.

For the treatment of defects in the coming sections, it will prove useful to introduce the following transformation to a parity symmetric basis
\bea
C_{\pm,l}^{\dagger} = \frac{e^{\pm {\rm i} \varphi}}{\sqrt{2}} \left(s_{\varphi,l}^{\dagger} \pm a_{\varphi,l}^{\dagger}\right),   \label{C_to_s_a}
\eea
where $\varphi$ is a phase, which is arbitrary for now. 
It is straight-forward to demonstrate that
\bea
H_{{\rm eff}} &=& H_s + H_a  
\nonumber \\ &=& 
 -t \sum_{l}^{} \left(s^{\dagger}_{\varphi,l}s^{\phantom\dagger}_{\varphi,l+1} + a^{\dagger}_{\varphi,l}a^{\phantom\dagger}_{\varphi,l+1} + h.c. \right), \label{zigzagH}
\eea
and that the creation operators on the carbon network are now given by
\bea
c_{j,l}^{\dagger} & = & \sqrt{\frac{2}{N}} \left( s_{\varphi,l}^{\dagger} \cos(\phi(l)+Qj-\varphi)
 \right. \\ \nonumber & & ~~~~~~~~~~~~~~~~~~~ \left. 
-{\rm i}~ a_{\varphi,l}^{\dagger} \sin(\phi(l)+Qj-\varphi) \right). \label{backtrafo}
\eea
Thus, $\varphi$ is nothing more than the phase of the wavefunction around the nanotube. 
Since the channels are degenerate, at this point the choice of $\varphi$ 
may seem arbitrary, but it is clear that the introduction of 
structural defects and impurities will in general lift this degeneracy. 
As will be illustrated below, in some cases the right choice of $\varphi$ allows for the decoupling of the $s_\varphi$ and $a_\varphi$ channels even in the presence of complicated
structural defects. 
Note, that even small perturbations e.g.~coming from the substrate will in general break the 
translational invariance around the tube, but not necessarily the reflection symmetry.
Therefore, in almost all physical situations
the symmetric/antisymmetric states with some choice of $\varphi$ will be the preferred 
basis.

\begin{figure}
\centering
(a) \includegraphics[width=0.44\textwidth]{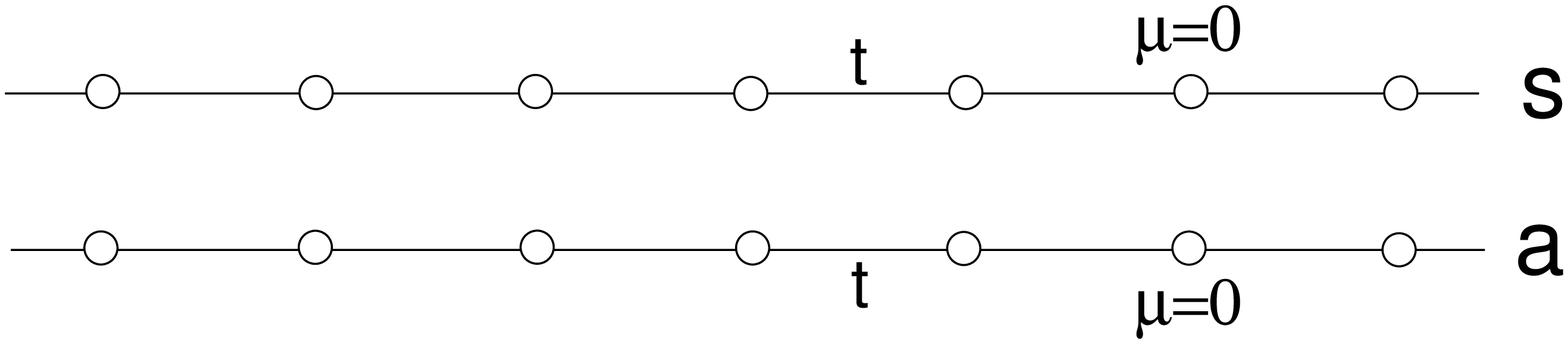}  \\
(b) \includegraphics[width=0.44\textwidth]{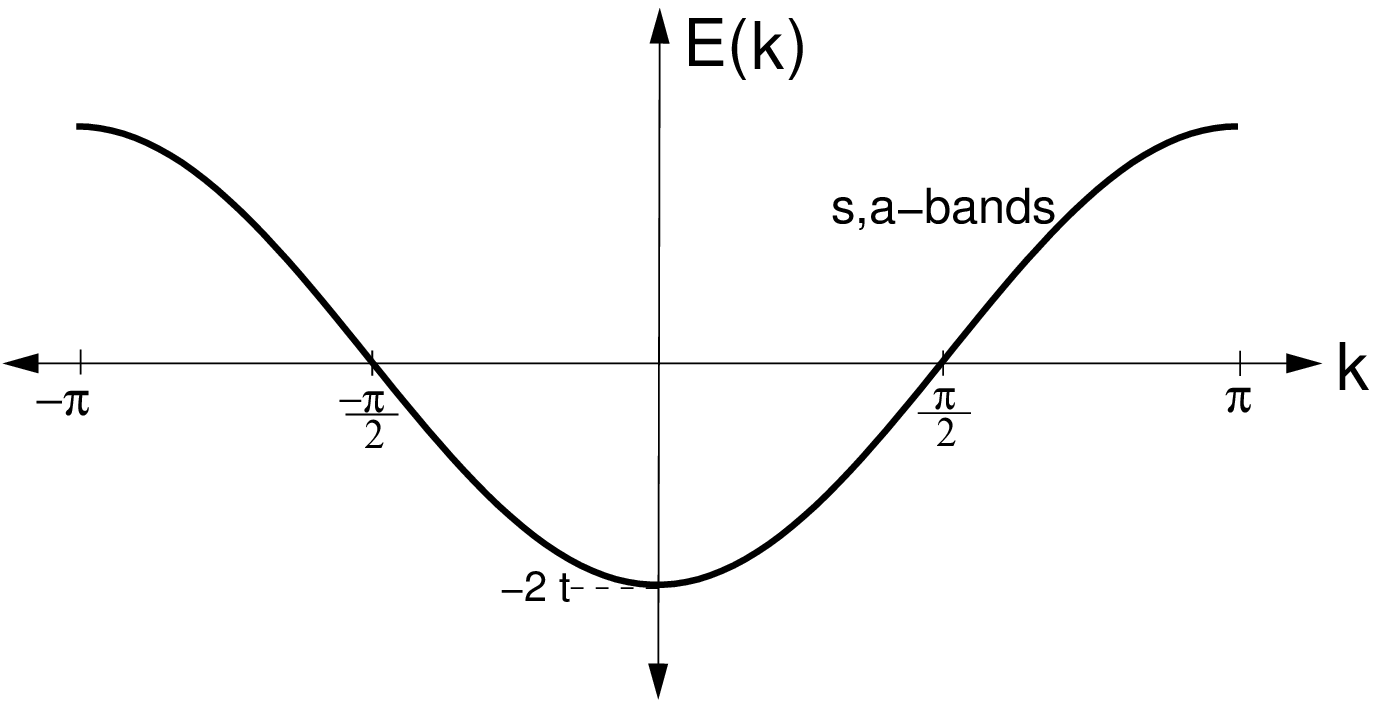}
\caption{(a) Effective model for the zigzag CNT. The hopping amplitudes are identical for both modes and at half filling $\mu=0$. (b) Since the effective model consists of two identical chains, the bands of the $s$ and $a$ channels lie on top of each other
(effective lattice spacing $a_{\rm eff}=a\sqrt{3}/4$). }\label{double_chain3}
\end{figure}

The effective lattice model for this basis 
and its band structure are shown in Fig.~\ref{double_chain3}.
Interestingly, the effective lattice model for the conducting channels has in fact a higher
translational symmetry than the original lattice.  The translation by
$\sqrt{3}a$ along the tube in the original lattice corresponds to
a translation by four sites in the effective chains,
so that the true wavevector along the tube is obtained only
after a four-fold zone folding in Fig.~\ref{double_chain3}(b).   \\

Rehybridization due to higher bands and curvature can also be taken into account
within this model for the metallic zigzag tubes.  
The corresponding hopping terms have been predicted within
the tight binding picture\cite{Kane,gap,kleiner} and result in a lower symmetry of the
model.  In particular, the effective model in Eq.~(\ref{zigzagH}) will also contain 
a very small alternating modulation of the hopping along each chain with a corresponding 
opening of a gap of a few meV.\cite{Kane,gap,kleiner}  
We can neglect this effect for excitations outside this energy range, which are in fact 
all but one or two for finite tubes of $L\alt 100nm$.
The degeneracy of the two channels is however {\it not} lifted by the curvature.

\subsection{Scanning tunneling microscopy}
The main goal of this study is to predict the effect on the electronic structure 
which can be very well measured by scanning tunneling spectroscopy (STS).
The corresponding STM images can be calculated from the simple models above in 
the following way:  First the eigenstates of finite chains are
determined depending on the lengths of the tubes, the 
boundary conditions, and possible defects, the description of which will be 
discussed in detail below.  
Secondly the resulting eigenstates are transformed 
back into the original basis of nanotube orbitals 
($\alpha_{x,y}, \beta_{x,y}$ or $c_{j,l}$)
as described in Eqs.~(\ref{H_arm}-\ref{backtrafo}).
Finally, the STM images for the eigenstates 
are calculated by standard assumptions of an s-wave orbital 
on the tip and $\pi$-orbitals for the nanotube.  For the images presented here 
we use the
parameters given in Ref.~[\onlinecite{meunier}] in arbitrary units for the tunneling
current.  In order to show the resulting impurity effects with a maximum 
resolution we choose a very small
distance $\alt \rm \AA$ of the STM tip to the sample. Larger distances naturally
result in plots with less resolution but look qualitatively the
same.  Orbital mixing and geometric distortion 
due to the nanotube curvature can also be taken into account,\cite{kleiner,meunier,deretz} but 
here the STM images are projected onto a flat surface in order to explicitly 
demonstrate the main effects of the defects.
The procedure above basically determines the LDOS of the nanotube by using the
square of the eigenstates in the effective models.

\begin{figure}
\centering
(a)~ \includegraphics[width=0.43\textwidth]{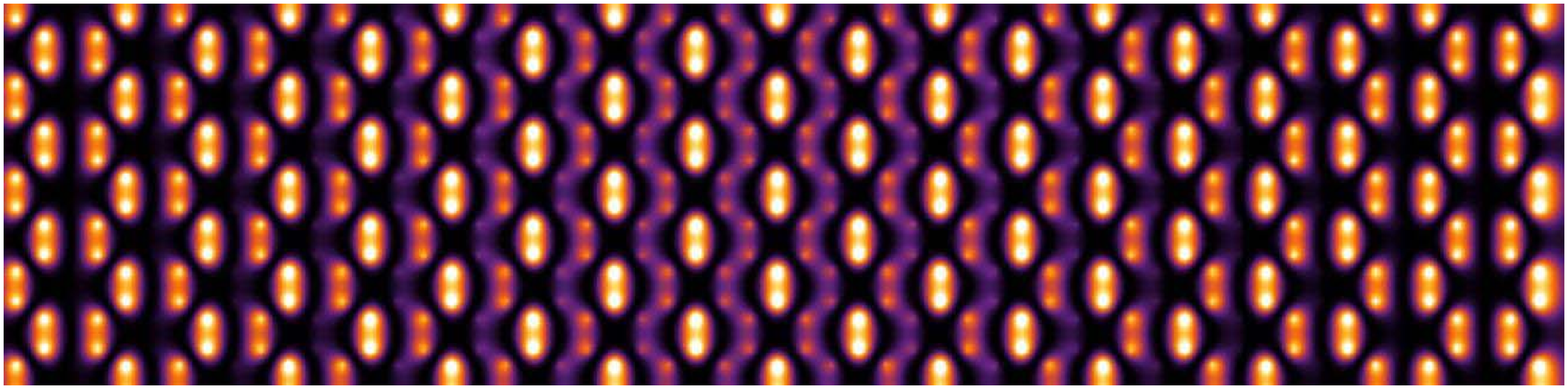}
(b)~ \includegraphics[width=0.43\textwidth]{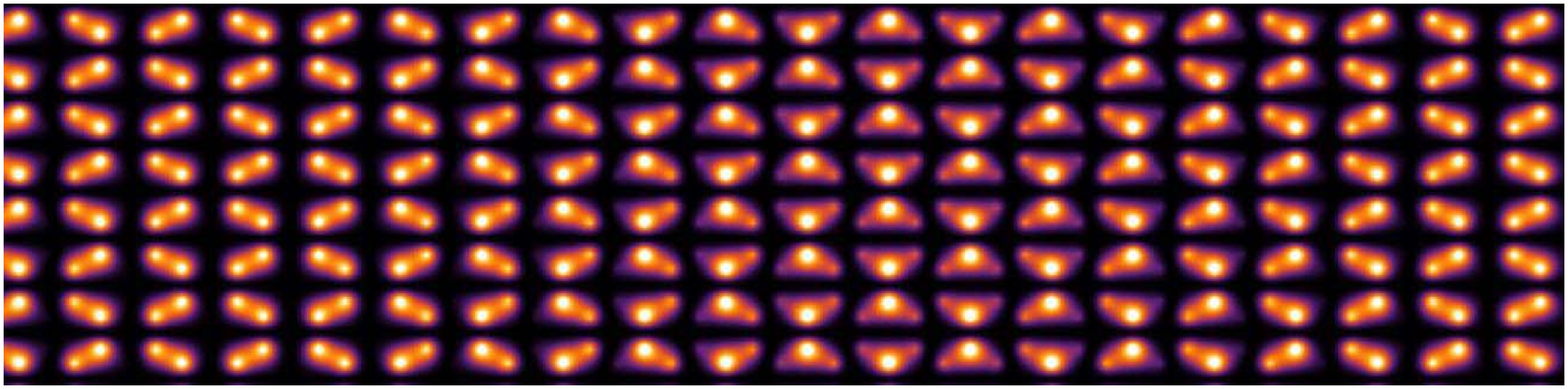}
(c)~ \includegraphics[width=0.43\textwidth,height=0.5cm]{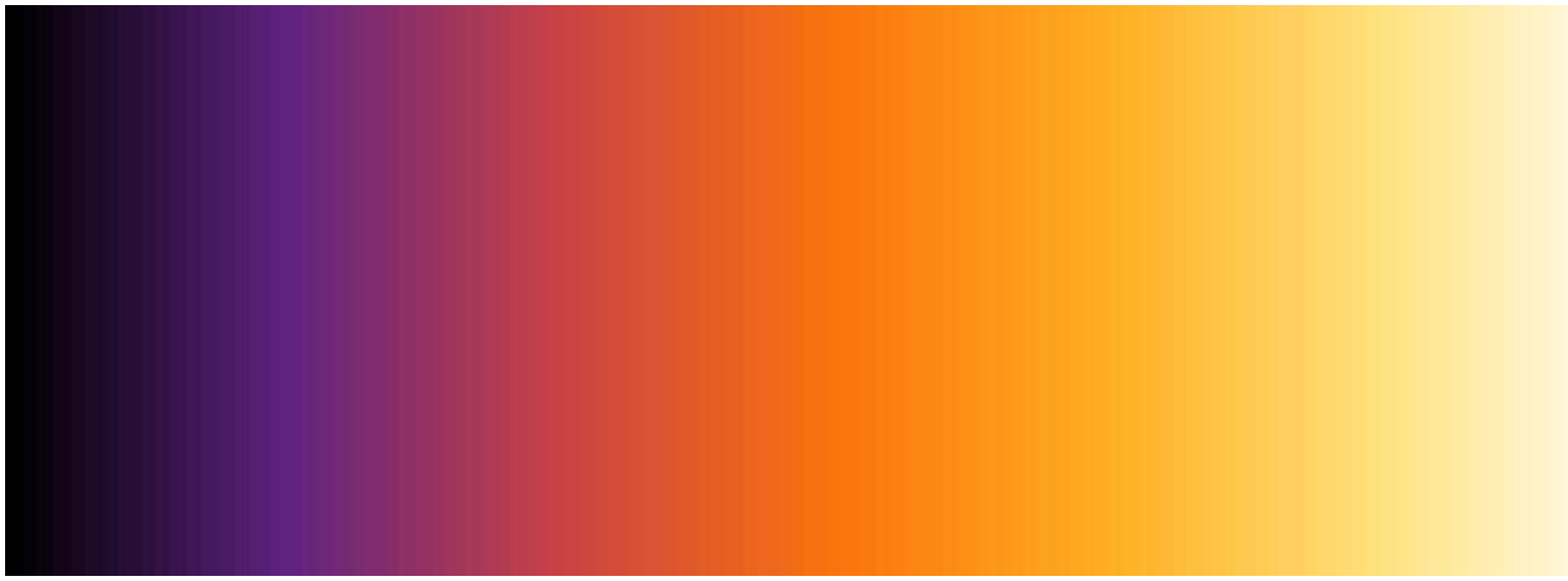}
\caption{(Color online) The expected STM image for an armchair tube with perfectly
cut edges.  A section near the right edge of a tube  of total length $30\rm nm$ is shown. 
(a) Image corresponding to the excitation of the symmetric mode ($S$) and
(b) of  the anti-symmetric channel ($A$).
(c) Color-scale of all STM images in arbitrary units from 0 to unity.}
\label{finitetube}
\end{figure}

In Fig.~\ref{finitetube} we show an example for a finite armchair tube of $30~{\rm nm}$
with sharp edges
which corresponds to 240 lattice sites in each chain of the effective model.
The symmetric mode for $k_y= \frac{162}{241}\pi$ ($E\approx 74\rm meV$)
and the antisymmetric mode for $k_y=\frac{159}{241}\pi$ ($E\approx 93 \rm meV$) are shown.
Here the energies are given relative to half-filling and 
we used hopping parameters which correspond to the accepted Fermi velocity
of $v\approx 8.1 \times 10^5\frac{m}{s}$.  The energy spacing for a finite tube of
$30~{\rm nm}$ is about $55~{\rm meV}$ for each mode (chain).
For the armchair tube the symmetric 
and antisymmetric modes are not degenerate.  
For the zigzag tube the two modes are degenerate,
but impurities and defects generally lift the degeneracy.
For example, in a 30nm
zigzag tube the cap described below in Sec.~\ref{IIIB} lifts the
degeneracies by $\Delta E \sim 10\rm meV$ whereas a single vacancy on the
same tube will have a much smaller effect giving $\Delta E \sim
\rm 0.1meV$.
The color scale given in Fig.~\ref{finitetube} has been used in all 
following STM images as well.

In the STM experiment the LDOS can be measured via
the differential conductance between the STM tip and the nanotube
\bea
\frac{dI}{dV}(eV,\vec{r}) \sim \sum_{|E_i-eV|<\delta} |\psi(E_i,\vec{r})|^2 ,
\label{sts}
\eea
where $\delta$ is the energy resolution of the measurement.
For short tubes individual states can be resolved, as was demonstrated
in several experiments.\cite{Venema,Lemay} 
For longer tubes it is still useful to consider individual eigenstates theoretically,
since the resulting experimental signal can always be constructed by a 
simple superposition of the nearly degenerate states.

The impurity effect on
the detailed discrete energy spectrum is not very easy to measure, while
the actual modification of the standing waves is much more visible, 
which we will discuss in this work.  
For the pure case as shown in Fig.~\ref{finitetube} two main features can be observed.
The short range feature on the atomic scale is characteristic for the 
detailed boundary condition as well as for the phase of the wavefunction.  
A long range feature
shows a rotation of an overall phase $\theta$ of the wavefunction,
which slowly changes the STM pattern along the 
chain.  This long-range feature is not related to impurity effects and only depends 
on the energy relative to the half-filled Fermi points.  In particular,
the rate of change with position 
is always given by $\theta \approx (k-k_F) \Delta y$, where $k-k_F \approx (E-E_F)/v$.

\begin{figure}
\centering
(a)~ \includegraphics[width=0.21\textwidth]{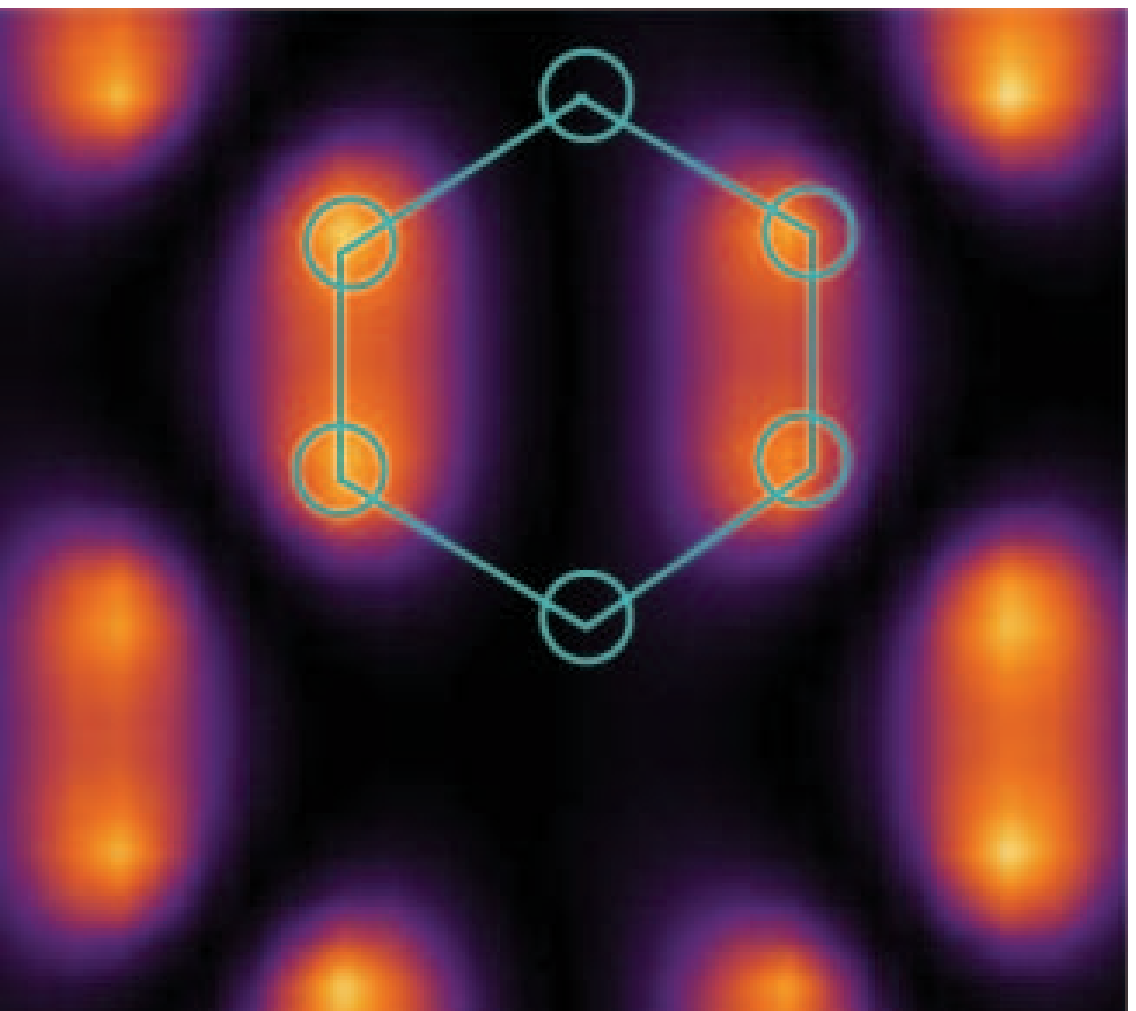}~ 
(b)~ \includegraphics[width=0.21\textwidth]{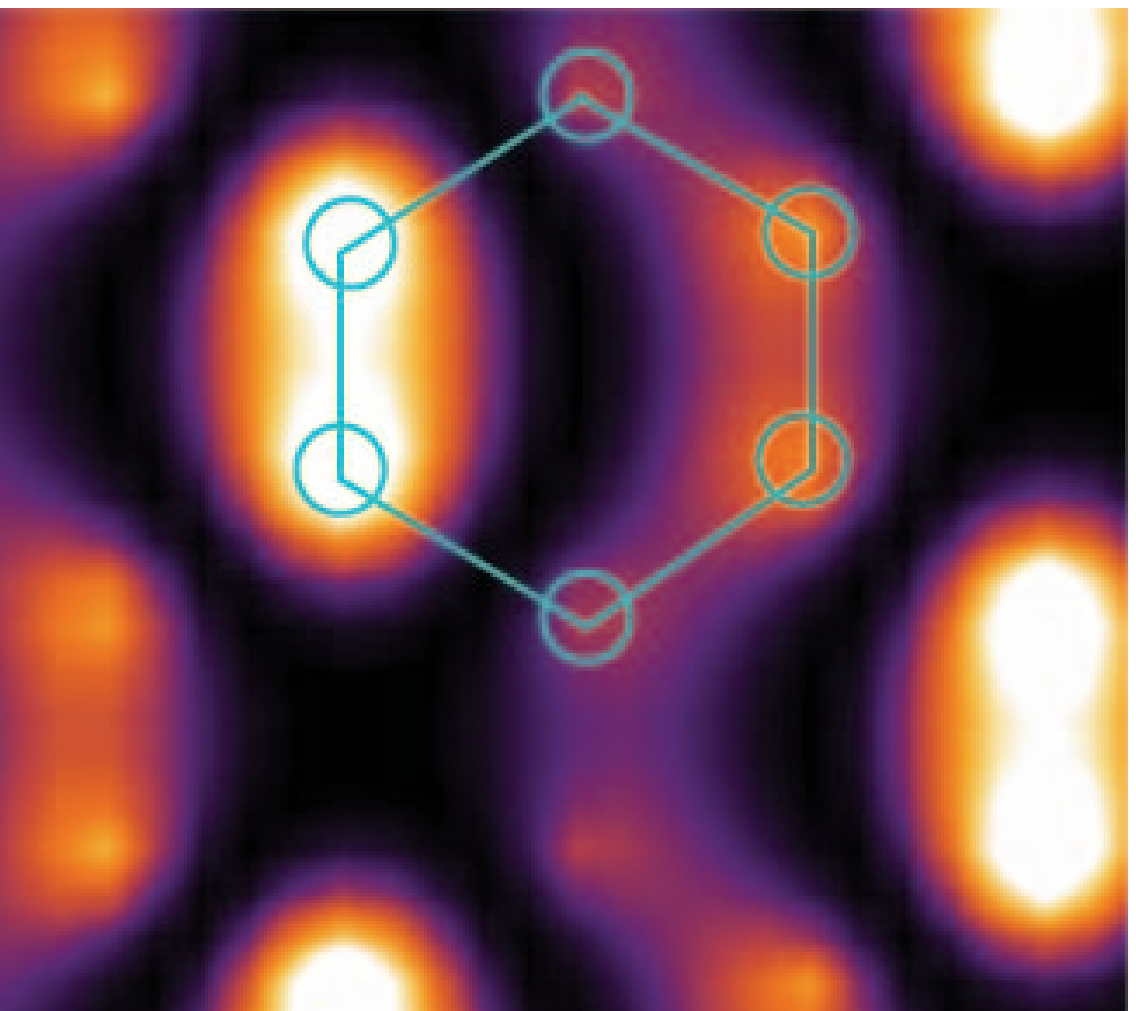}  \\
(c)~ \includegraphics[width=0.21\textwidth]{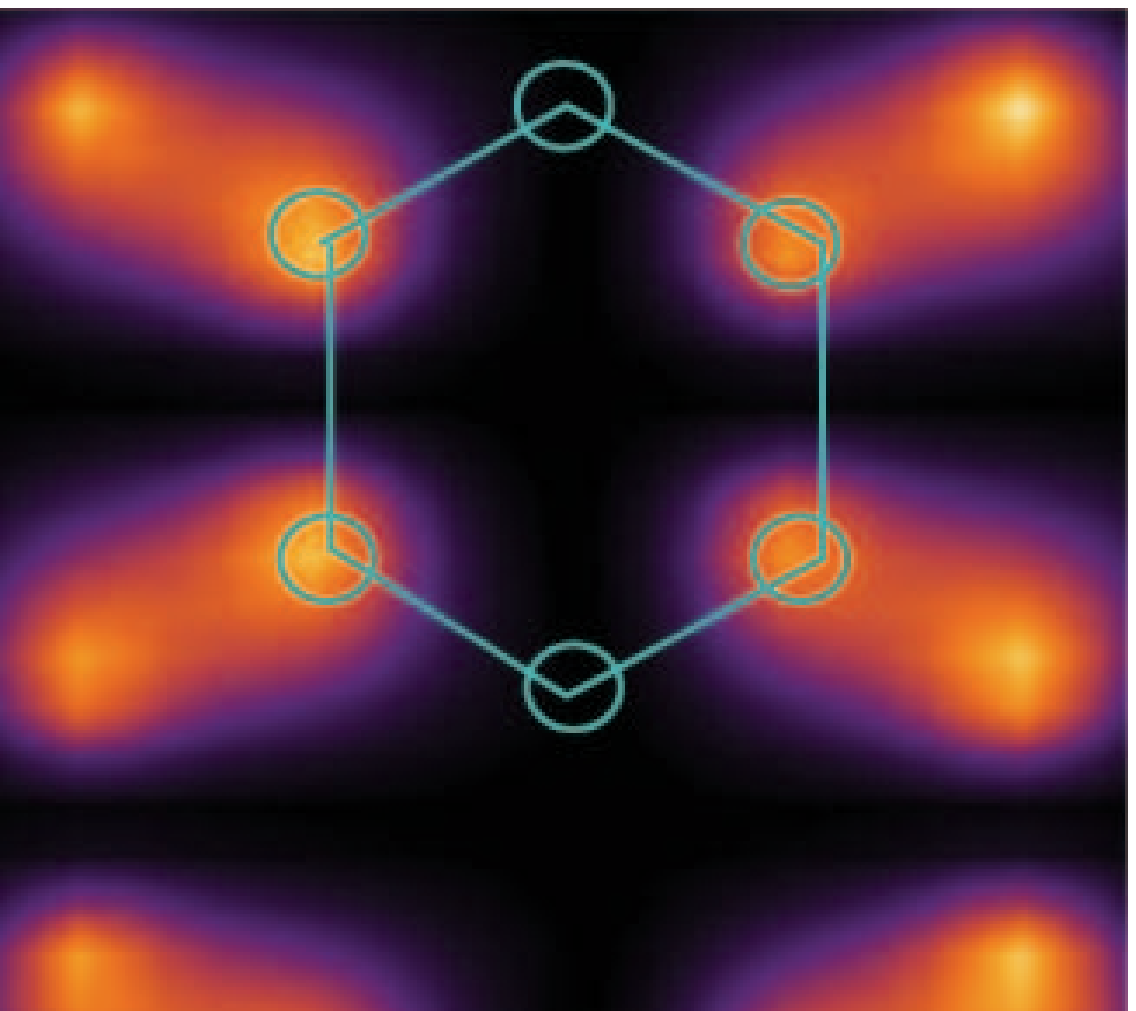}~ 
(d)~ \includegraphics[width=0.21\textwidth]{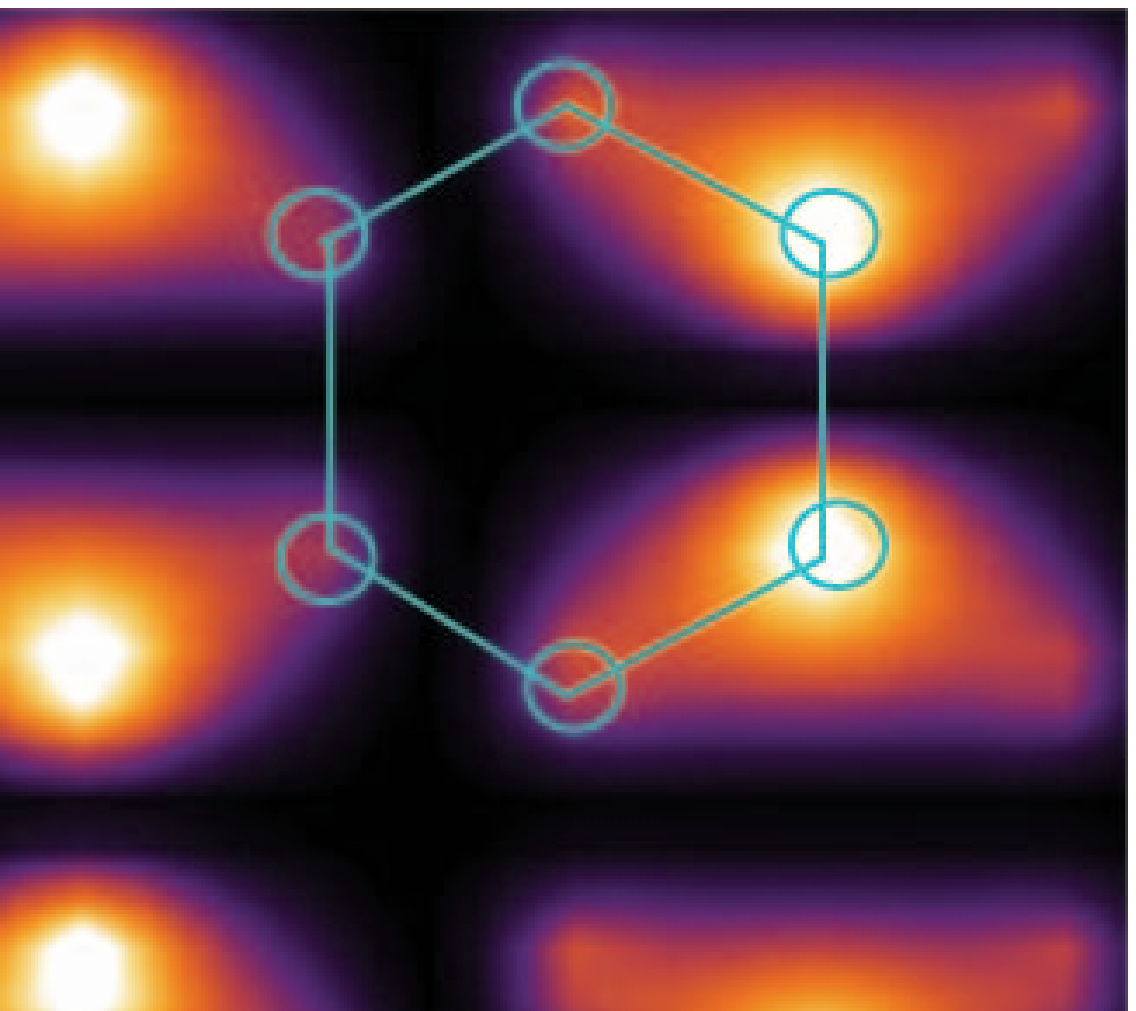}
\caption{(Color online) ARMCHAIR: Expected density patterns from an STS experiment.
The axis of the tube is the horizontal direction. (a) Image corresponding to the excitation of the symmetric mode ($S$) close to the edge of the nanotube. (b) Excitation of (a) is now phase shifted by  $\theta=\pi/2$. (c-d) Same as (a-b) but now for the anti-symmetric channel ($A$).}\label{STM_perfect_ARM}
\end{figure}

The short range STM pattern on the other hand is characteristic for each mode
(symmetric or anti-symmetric or a mixture).  It is also an indication for the 
detailed boundary condition close to the defects.  
Typical density patterns for the symmetric and antisymmetric states are shown in
Fig.~\ref{STM_perfect_ARM} for an armchair tube (enlargement from Fig.~\ref{finitetube}).  
The symmetric states show significant spectral weight on the 
circumferential bonds, while the antisymmetric states always have nodes there.
The short range patterns are useful to classify the symmetry of the states and the overall 
phase, which only changes slowly along the chain as described above.
Accordingly, the analysis of STM patterns close to 
boundaries and defects can be analyzed to classify the 
nature and symmetry of the perturbation as we will see later.

\begin{figure}
\centering
(a)~ \includegraphics[width=0.21\textwidth]{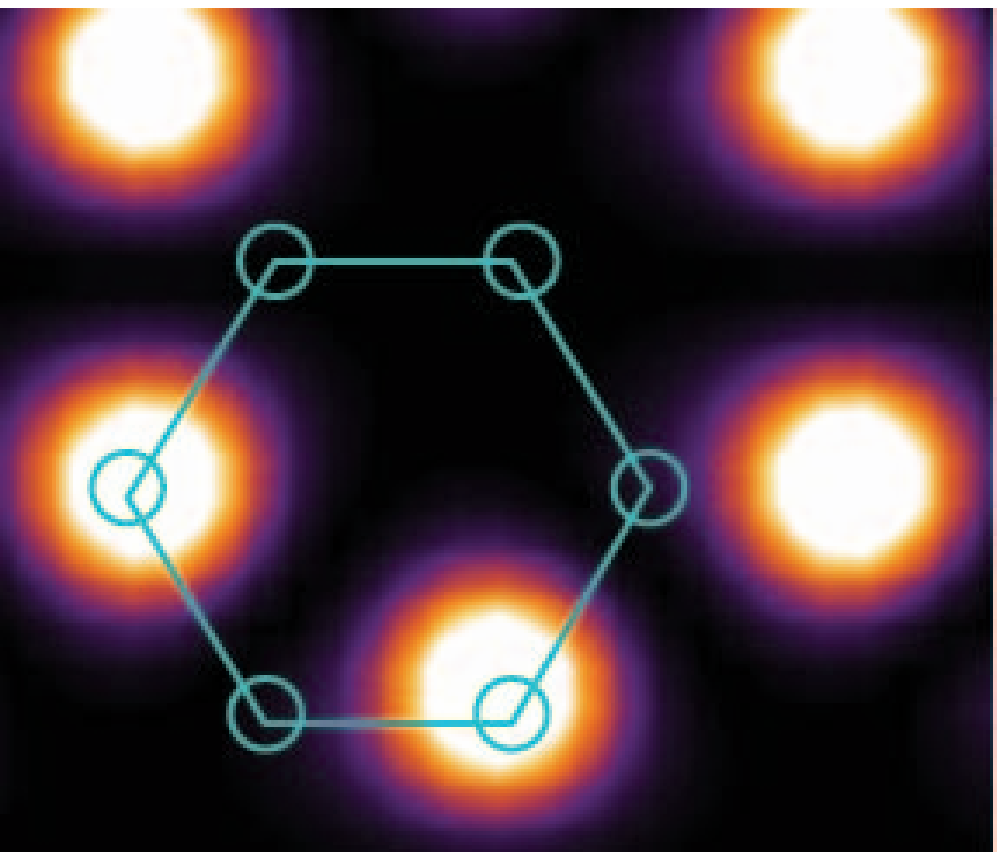}~ 
(b)~ \includegraphics[width=0.21\textwidth]{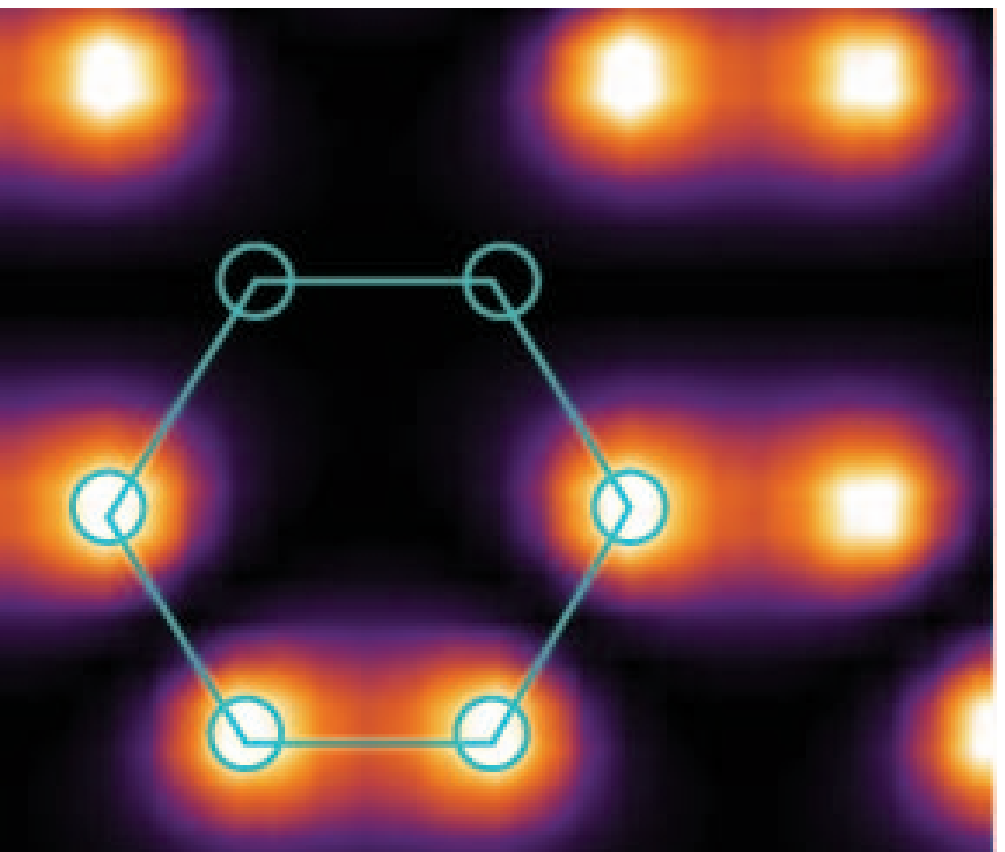}  \\
(c)~ \includegraphics[width=0.21\textwidth]{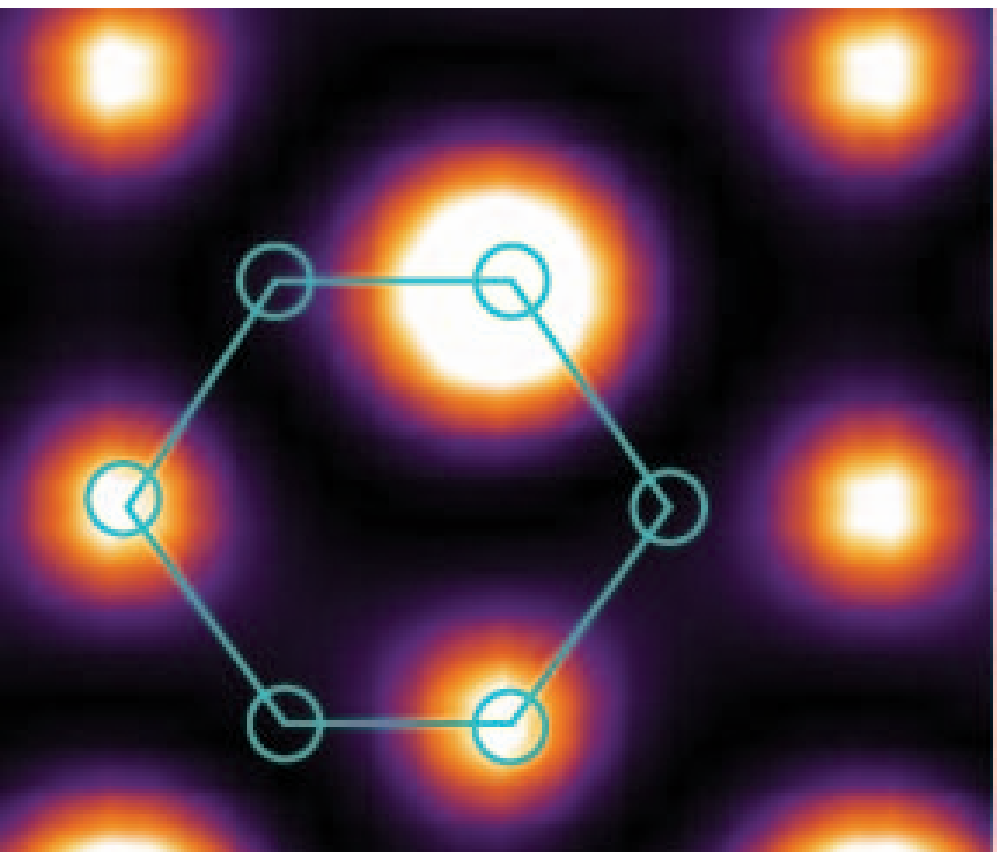}~ 
(d)~ \includegraphics[width=0.21\textwidth]{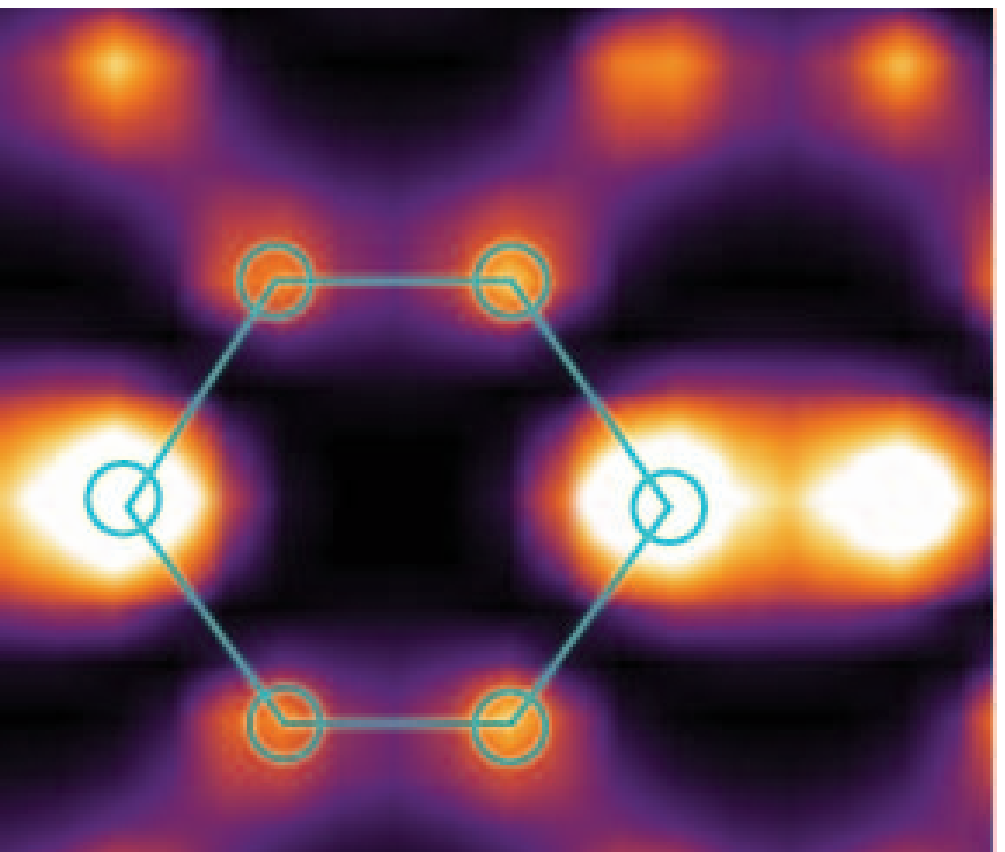}  \\
\caption{(Color online) ZIGZAG: Density patterns for the degenerate $s$ and $a$ channels near $E_F$. The phase is chosen to be $\varphi=0$ and the tube axis is aligned in the horizontal direction. (a) Image corresponding to the excitation of the $s_0$-mode close to a perfect
 end of the tube. (b) The state in (a) is now dephased by $\theta=\pi/2$ along the tube. (c-d) Plots analogous to (a-b) but now for the $a_0$-channel.} \label{STM_perfect_ZZ}
\end{figure}

For the zigzag tube the calculation of the STM images proceeds analogously.
However, in this case the symmetric and antisymmetric 
states for the pure case are degenerate.  For most impurities 
this degeneracy is lifted while still preserving the symmetry.
However, 
the choice of the phase $\varphi$ in Eqs.~(\ref{C_to_s_a}) 
and (\ref{backtrafo}) is no longer 
arbitrary and depends on the detailed impurity model.
Therefore, the analysis of the short range STM patterns for the zigzag tube
gives important additional information.
The characteristic short range patterns of a finite tube are
shown in Fig.~\ref{STM_perfect_ZZ}  for 
$s_{\varphi=0}$ and $a_{\varphi=0}$.  
Other possible values of $\varphi$ will be discussed in the context
of defects.

\section{CNT boundaries}  \label{boundaries}

In this section we examine the consequences of structures at the ends of SWCNTs
on their electronic properties. The tubes can generally have different kinds of
non-perfect  edges such as extra atoms at open ends, pentagons, or half fullerene caps
that can crucially modify their behavior.
In what follows we study this problem in the framework of the chain lattice
models derived in Sec.~\ref{Eff_Mod}. This  will illustrate how simple imperfect ends or even
complex carbon structures that close the tube (caps), can be easily incorporated
to the effective Hamiltonians by adding only one or a few sites to the initial
two chain models representing the clean nanotube.

\subsection{Armchair}
\begin{figure}[tp]
\centering
(a) \includegraphics[angle=270,width=0.2\textwidth]{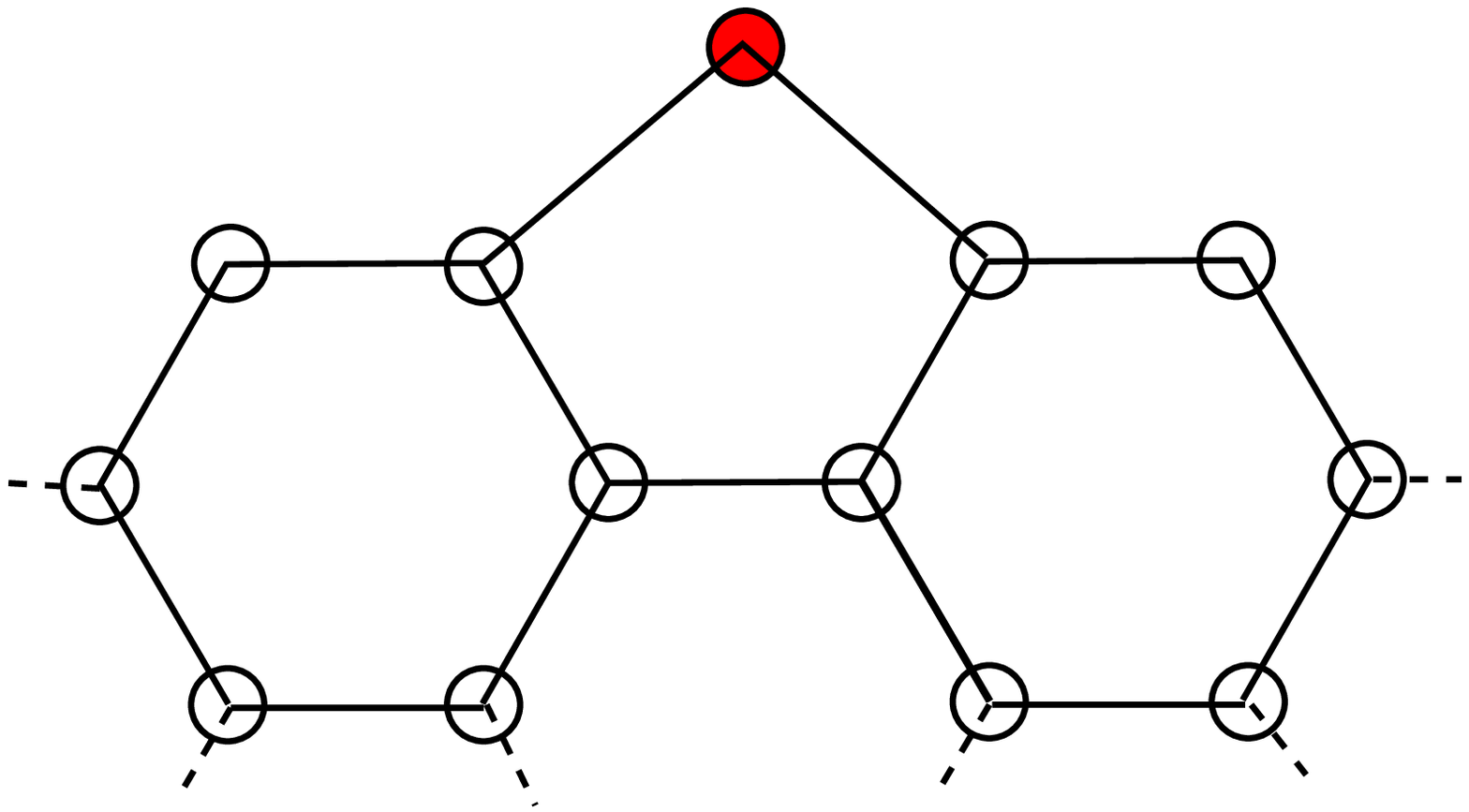}
~~(b) \includegraphics[angle=270,width=0.2\textwidth]{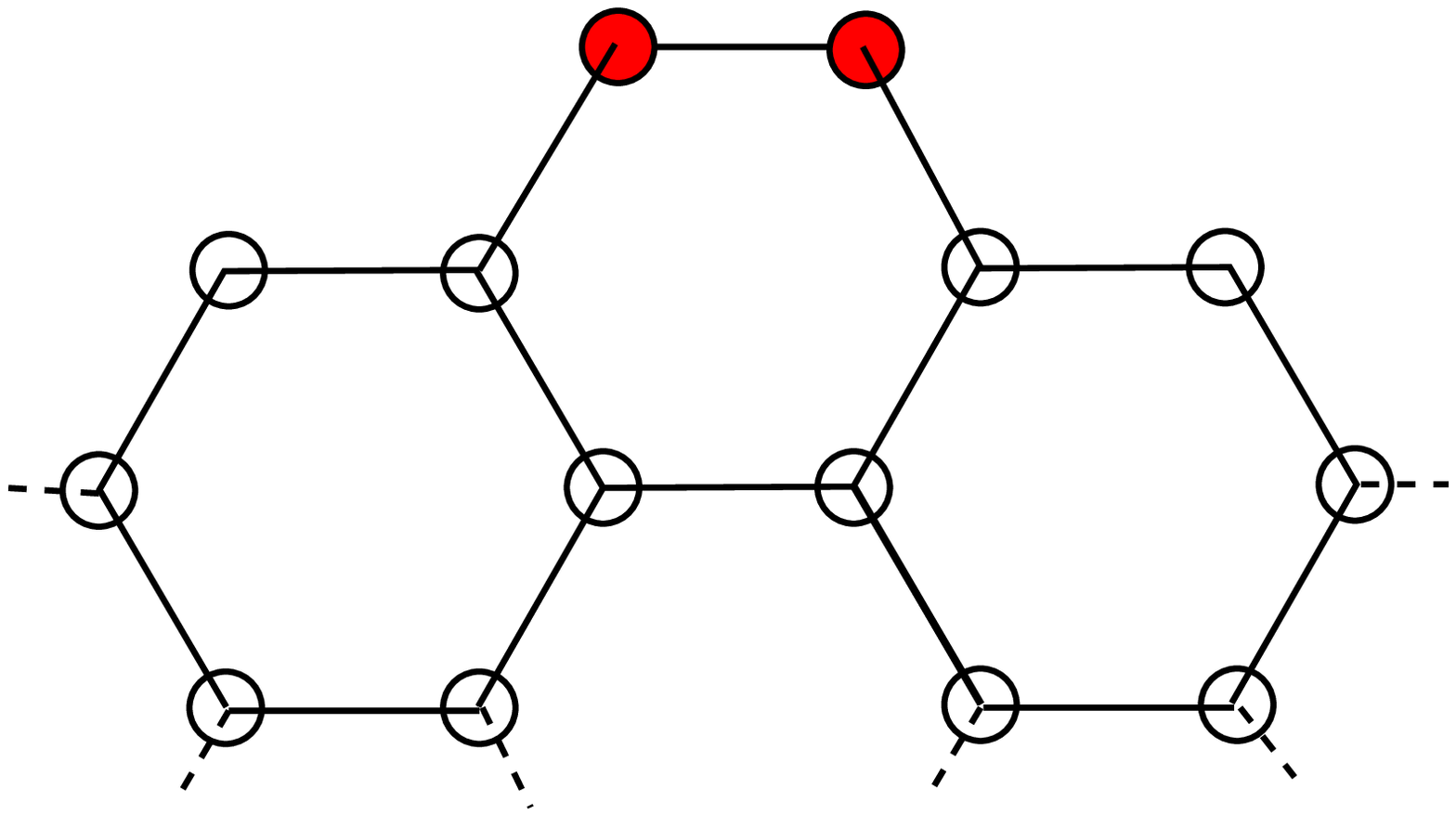}
\caption{ (Color online) ARMCHAIR: (a) Added carbon atom at the end of the tube. (b) Extra dimer at the edge. }\label{ARM_boundary_def}
\end{figure}

The simplest non-trivial edge that can be introduced
is the addition of a single carbon atom
at an otherwise perfect open end (clean cut)
of a SWCNT as shown in  Fig.~\ref{ARM_boundary_def}(a).
The extra site can be included in
the full hopping Hamiltonian of an armchair tube in Eq.~(\ref{H_arm}) by an additional term
\bea
H_C = -t \left( \beta^{\dagger}_{x=1, y=L} + \alpha^{\dagger}_{x=2, y=L} \right) c + h. c.,
\eea
 where $c$ represents the site of the additional carbon atom.
Since the conduction bands only contain the $k=0$ modes, using the
approximation in Eq.~(\ref{FT_armchair}) the additional term simplifies to
\bea
H_{C_{\rm eff}} &=& -\frac{t}{\sqrt{N}} \left( (\beta^{\dagger}_{y=L} + \alpha^{\dagger}_{y=L})  c + h. c. \right) \label{xtra_arm} 
\\  \nonumber &=& 
 -\frac{t}{\sqrt{N}} \left( S^{\dagger}_{L} c + h. c. \right).
\eea
Therefore, the electron on the extra site $c$ hops to the symmetric
$S$-mode chain, while the antisymmetric mode is unmodified by the
additional carbon atom as would be expected from the geometry of the problem.
The effective hopping to the impurity weakens as the radius of the tube increases
($t'=t/\sqrt{N}$) since the $S$-mode 
is distributed homogeneously around the tube of which 
only two bonds couple to $c$.

Another possible defect is the presence of an extra bond as shown in Fig.~\ref{ARM_boundary_def}(b). In a similar way as above for the additional atom, 
the extra bond (B) can be included into the effective model by adding
\bea
H_{B_{\rm eff}} & =&  -\frac{t}{\sqrt{N}} (\beta^{\dagger}_{y=L} \alpha^{\phantom\dagger}_{B} + \alpha^{\dagger}_{y=L} \beta^{\phantom\dagger}_{B}+h.c.) 
\\ & & ~~~~~~~~~~~~~~~~~~  \nonumber
-t ( \alpha^{\dagger}_{B} \beta^{\phantom\dagger}_{B} + h.c.)   \nonumber \\
& =&  -\frac{t}{\sqrt{N}} \left( (S^{\dagger}_{L} S^{\phantom\dagger}_B - A^{\dagger}_{L} A^{\phantom\dagger}_B) + h. c. \right) 
\\ & & ~~~~~~~~~~~~~~~~~~  \nonumber
- t (S^{\dagger}_{B} S^{\phantom\dagger}_B  -  A^{\dagger}_{B} A^{\phantom\dagger}_B).   
\eea
As a result the defect is incorporated by adding an extra site to {\it both} $S$ and $A$ modes with a weakened hopping amplitude $t'=t/\sqrt{N}$. It can be easily checked that by systematically adding all possible bonds around the tube ({\it i.e.}~extending it), the resulting modification to the effective Hamiltonian will be a mere added site, as it should be.

It is worth mentioning that none of the two defects considered so far connects the pair of chains of the effective model. An example of such mixing produced by a defect will be provided later in Sec.~\ref{surf_def_SW_arm}.
Therefore, the modification of the STS spectrum by the extra sites
can be determined in a straight-forward way: (i) The extra carbon atom will have a finite 
spectral density only when a state corresponding to $S$ is excited; otherwise it will be completely empty. (ii) STS measurements performed over an extra bond are expected to yield the same density as the one measured over the bonds at the edge of a perfect nanotube that is extended by $a/2$.  
Otherwise the pattern follows the pure case in Fig.~\ref{STM_perfect_ARM} 
with only a slight shift 
of the phase $\theta$.
\\

\begin{figure}[tp]
\centering
(a)~\includegraphics[width=0.3\textwidth]{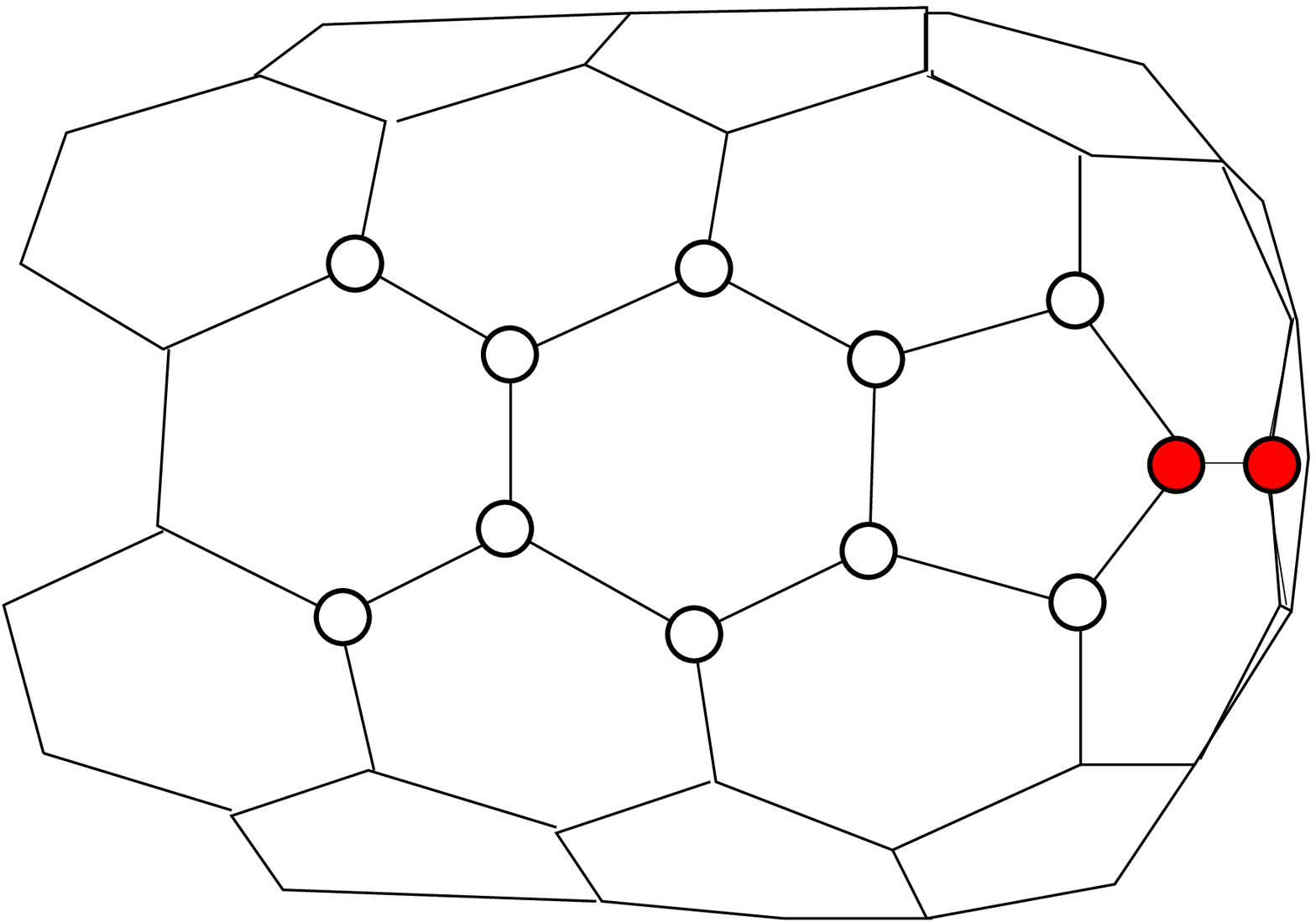} 
~ (b)~\includegraphics[width=0.44\textwidth]{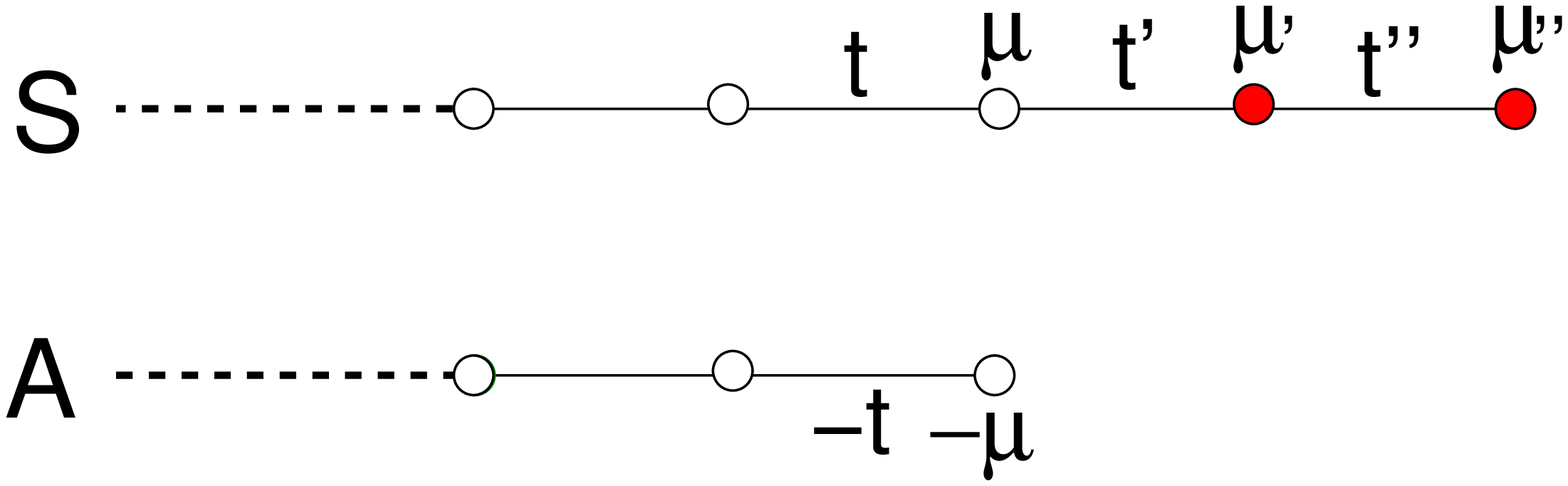}  \\
(c)~\includegraphics[width=0.44\textwidth]{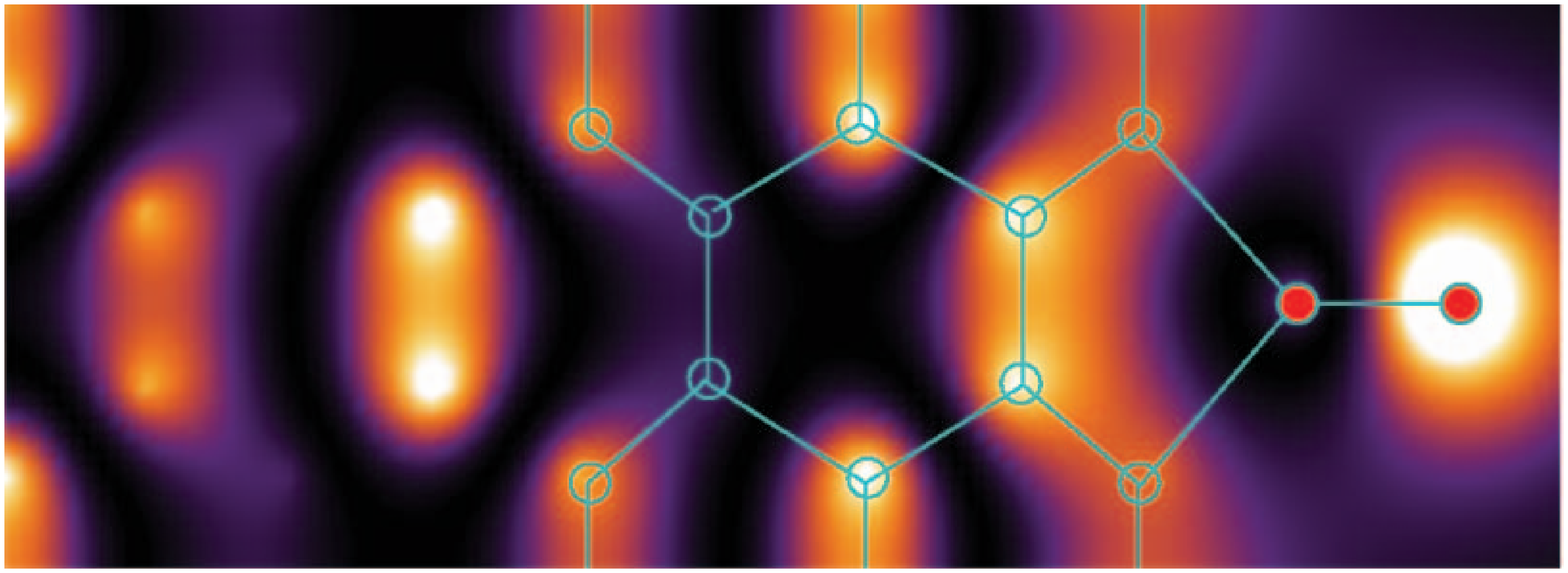}  \\
(d)~\includegraphics[width=0.44\textwidth]{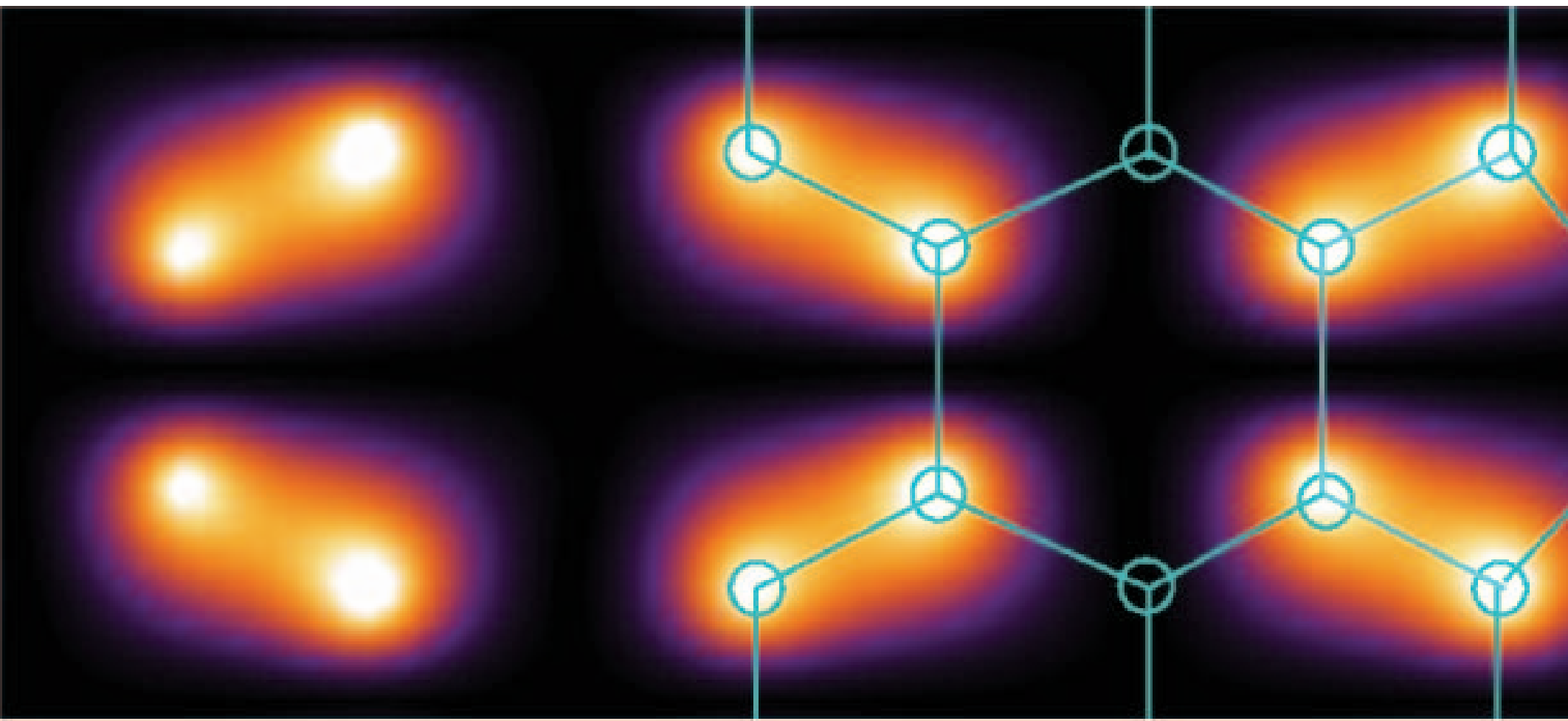}
\caption{(Color online) ARMCHAIR: (a) Half fullerene cap at the end of a $(6, 6)$ carbon nanotube. (b) Low energy effective model for the $(6, 6)$ nanotube including a half fullerene cap. The sites corresponding to the cap are shown in red (gray). The $A$-mode is not modified by the cap. Parameters for the added sites are: $t'=\sqrt{2}t$, $\mu'=0$, $t''=t$, $\mu''=2t$. (c) Electronic density for an eigenstate of the symmetric ($S$) mode, obtained using the effective model in (b). For illustration purposes sites corresponding to the cap have been flattened out. (d) Same as in (c) but now for the $A$-mode; as expected, the cap is empty in this case.}\label{pic_cap_arm}
\end{figure}

SWCNTs may also be closed at their ends by some carbon structure such as a half fullerene or a nanocone. The closure is made possible by the introduction of topological defects
(typically pentagons)
which induce a bending in the carbon network.
For the armchair CNT we focus here on the $(6, 6)$ nanotube closed by a half fullerene.
The corresponding cap introduces six pentagon defects in order to close the tube,
as it can be seen in Fig.~\ref{pic_cap_arm}(a).

To add the caps to the effective model, the only necessary modification is the
inclusion of two extra sites at each end of the $S$-mode chain. The first one of them
represents the $k=0$ mode of the ring formed by the atoms at the tip of the $6$ pentagon defects near the end as shown in Fig.~\ref{pic_cap_arm}(a). The last site corresponds to the $k=0$ mode of the hexagon at the top of the cap.
The values of the hopping and chemical potential for each one of these extra 
sites will be different from the bulk values $t'=\sqrt{2}t$, $\mu'=0$, $t''=t$, $\mu''=2t$
as indicated in Fig.~\ref{pic_cap_arm}.
Due to the pentagon defects of the cap, the chain corresponding to the antisymmetric mode is not modified by the addition of this structures at the ends of the tube. It is easy to understand this by considering that the atoms at the tips of the pentagon defects are added in the same way as explained in Eq.~(\ref{xtra_arm}) for a single pentagon. 
In this picture the $A$-mode becomes then completely decoupled from the half fullerene cap.\cite{loc_state} 
As a result, the effective lattice model that includes the cap looks like the one depicted in Fig.~\ref{pic_cap_arm}(b).
Although we have here
outlined the procedure for the specific cap shown in Fig. \ref{pic_cap_arm}(a),
the same basic ideas can be applied to construct effective models for
nanotubes of larger radius. Capping structures for such tubes will in
general be composed by a larger number of carbon atoms; therefore, the
resulting effective models will require more additional sites for its
description.

For the individual states near a capped nanotube
very clear effects in an STS experiment are expected. Namely, if the
applied voltage matches the energy of one of the states corresponding
to the $A$-mode, the cap will appear completely empty. On the other hand,
exciting a symmetric state would be clearly recognizable by the appearance
of a finite amount of electron density distributed over the cap. Our density
maps (simulated STM images) clearly illustrate this effect as shown in
Fig.~\ref{pic_cap_arm}(c) and (d). 
The density over the body of the tube is 
distributed very similar to the pure case in Fig.~\ref{STM_perfect_ARM}.

\subsection{Zigzag} \label{IIIB}
As already mentioned in Sec.~\ref{Eff_Mod} B there is an arbitrary phase $\varphi$ that defines the transformation to the effective two chain models. In what follows we will choose $\varphi$ for each structural defect that we treat in such a way that their corresponding low energy models are as simple as possible.

\begin{figure}[tp]
\centering
(a)~\includegraphics[angle=270,width=0.15\textwidth]{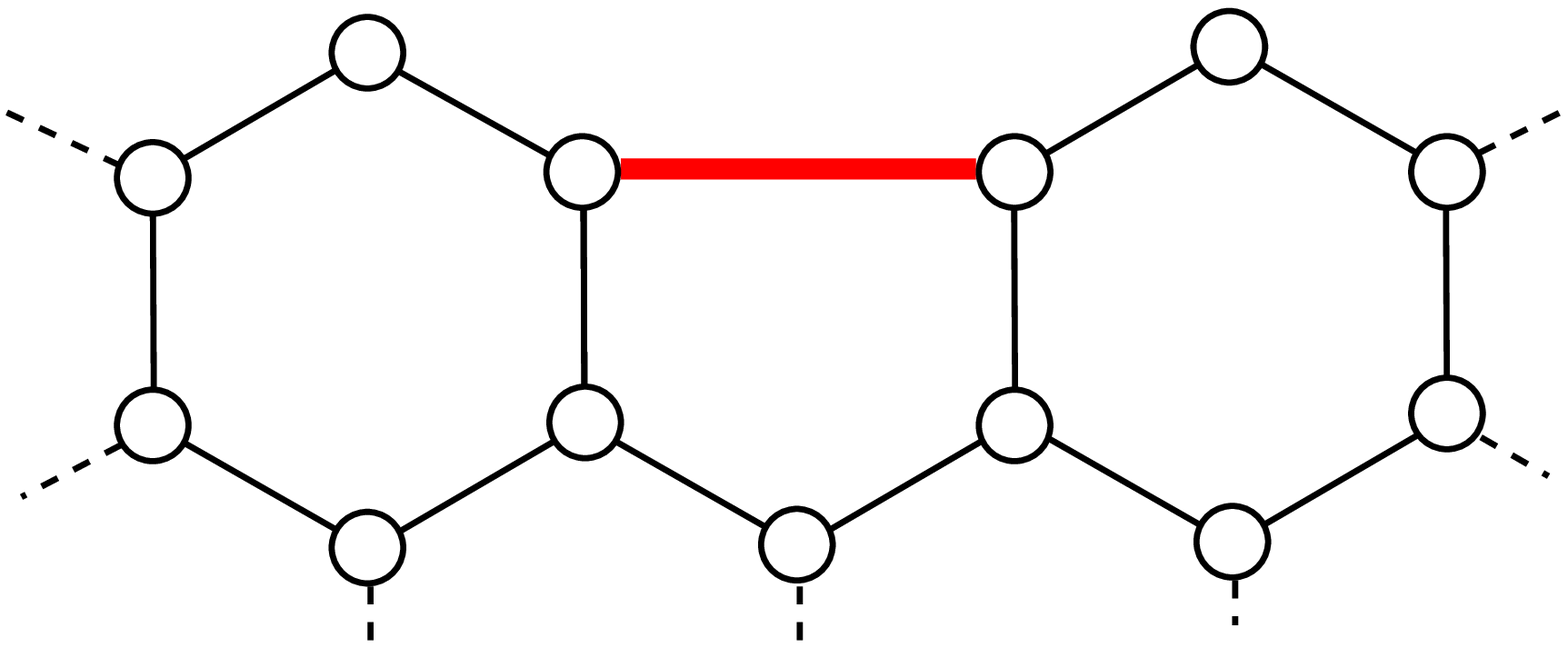}
~~ (b)~\includegraphics[angle=270,width=0.2\textwidth]{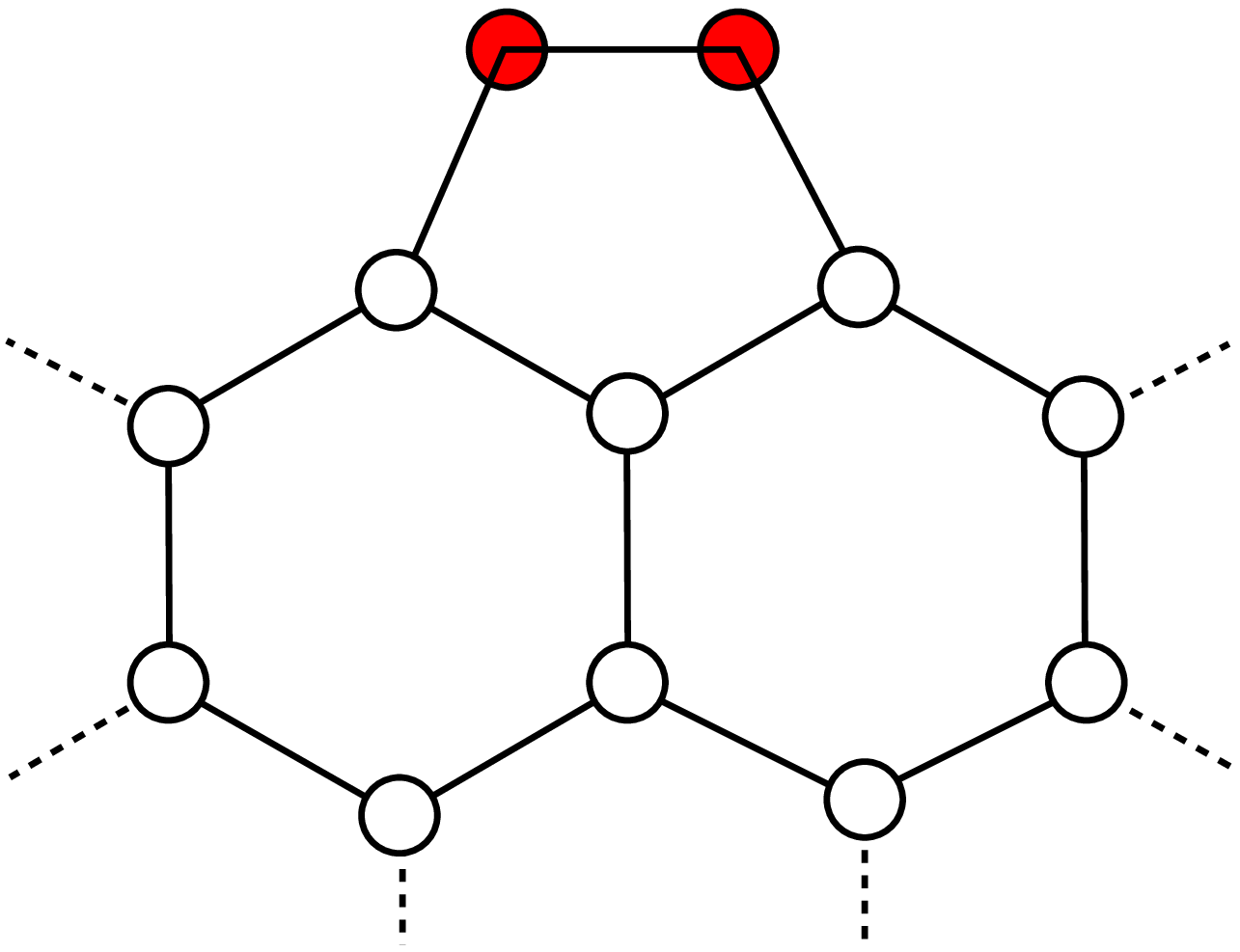}
\caption{(Color online) ZIGZAG: (a) A vacancy at the edge of the tube. (b) Extra dimer producing a pentagon defect at one end of the CNT. }\label{ZZ_boundary_def}
\end{figure}

We first consider the absence of a
carbon atom at the end as illustrated in  Fig.~\ref{ZZ_boundary_def}(a).
This sort of imperfection involves the elimination of a pair of bonds that joined the missing atom at site $l=1$ with a pair of atoms at $l=2$ plus the addition of an extra hopping matrix between them. The resulting effective impurity Hamiltonian for such a vacancy (V) is
\bea
H_{V_{\rm eff}} & =&  \frac{2t}{N} \left( s^{\dagger}_{\frac{\pi}{6},1}s^{\phantom\dagger}_{\frac{\pi}{6},2}   + h.c. \right) 
\nonumber \\ & & ~~~~ 
- \frac{t}{N} \left( s^{\dagger}_{\frac{\pi}{6},2}s^{\phantom\dagger}_{\frac{\pi}{6},2} - 3 a^{\dagger}_{\frac{\pi}{6},2}a^{\phantom\dagger}_{\frac{\pi}{6},2}  \right) , 
\eea
where we have chosen $\varphi=\pi/6$. The first term corresponds to the elimination of the atom and the second to the inclusion of the extra bond shown in Fig.~\ref{ZZ_boundary_def}(a).

Another possible defect is the addition of an extra bond as the one shown in Fig.~\ref{ZZ_boundary_def}(b). In the effective models it is represented by couplings
of both chains to the bond sites, corresponding to the impurity Hamiltonian 
\bea
H_{B_{\rm eff}} & = & t \left( b^{\dagger}_{-} b^{\phantom\dagger}_{-} - b^{\dagger}_{+} b^{\phantom\dagger}_{+} \right) 
\nonumber \\ & &  
-\frac{t}{\sqrt{N}} \left( \sqrt{3} s^{\dagger}_{0,1}b^{\phantom\dagger}_{-} - {\rm i}~ a^{\dagger}_{0,1}b^{\phantom\dagger}_{+} + h.c. \right) , 
\eea
where,
\bea
b^{\dagger}_{\pm} = \frac{b^{\dagger}_{1} \pm b^{\dagger}_{2}}{\sqrt{2}} , \label{bond_trafo}
\eea
and $b_{1,2}$ are the sites of the extra bond. We note that the choice $\varphi=0$ keeps the two modes completely decoupled even in the presence of a bond defect at the edge.  
Therefore, the STS signal will look very similar as in the pure case in Fig.~\ref{STM_perfect_ZZ} (except for a small change of the phase $\theta$ {\it along} the chain).

\begin{figure}[tp]
\centering
(a)~\includegraphics[width=0.3\textwidth]{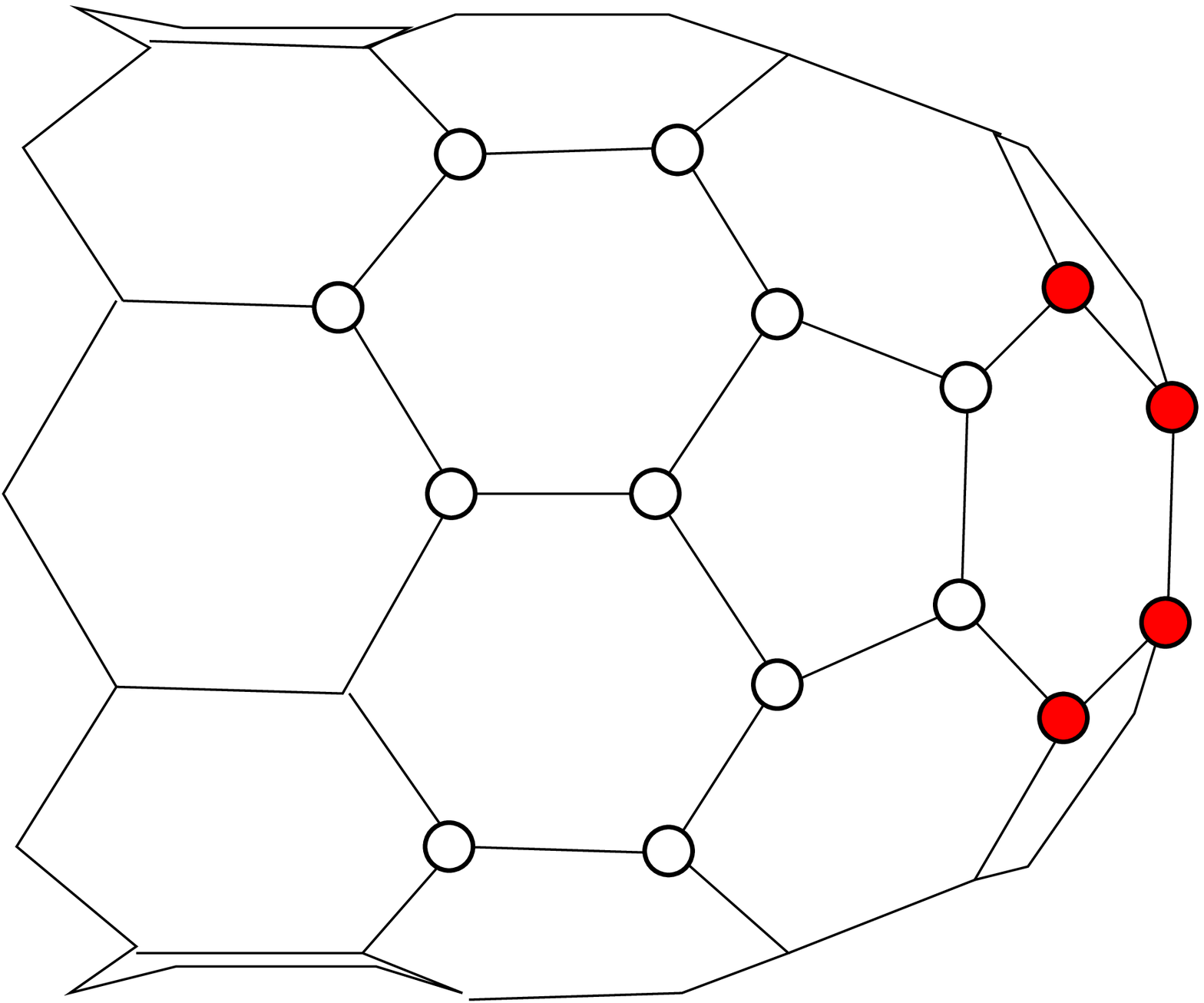} ~ (b)~\includegraphics[width=0.44\textwidth]{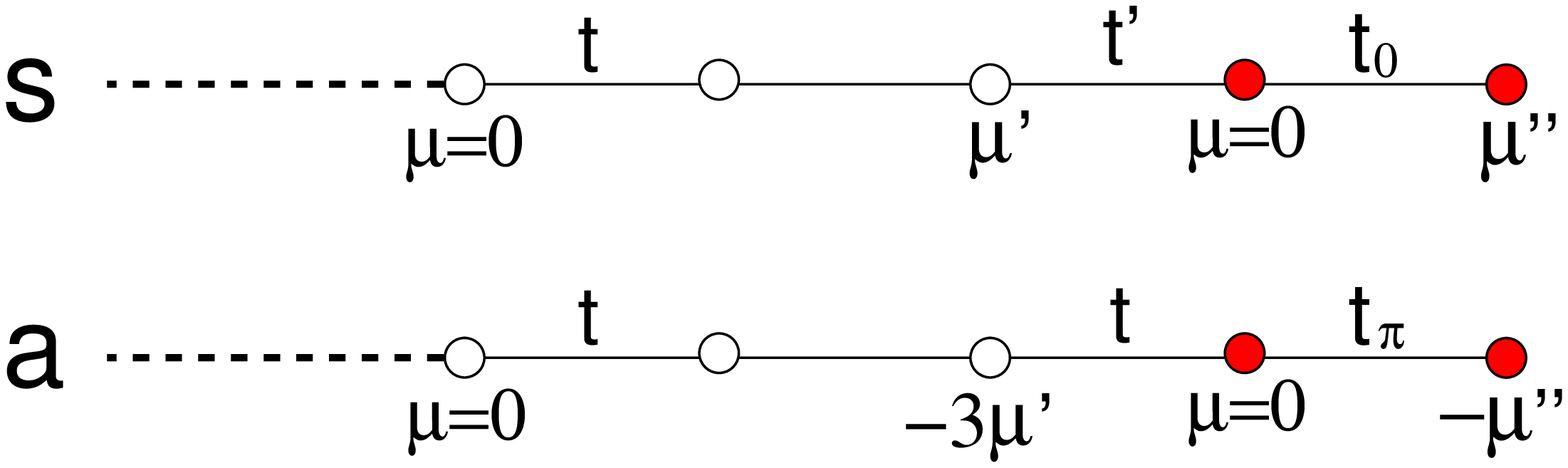}  \\
(c)~ \includegraphics[width=0.44\textwidth]{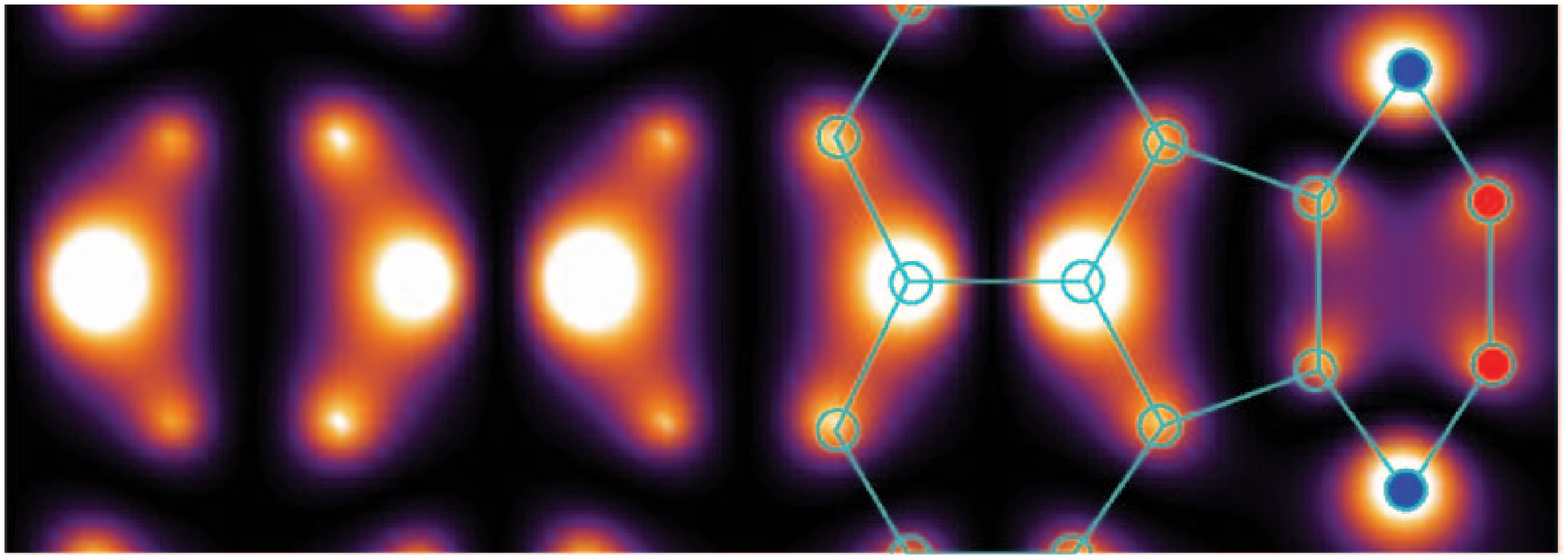}  \\
(d)~ \includegraphics[width=0.44\textwidth]{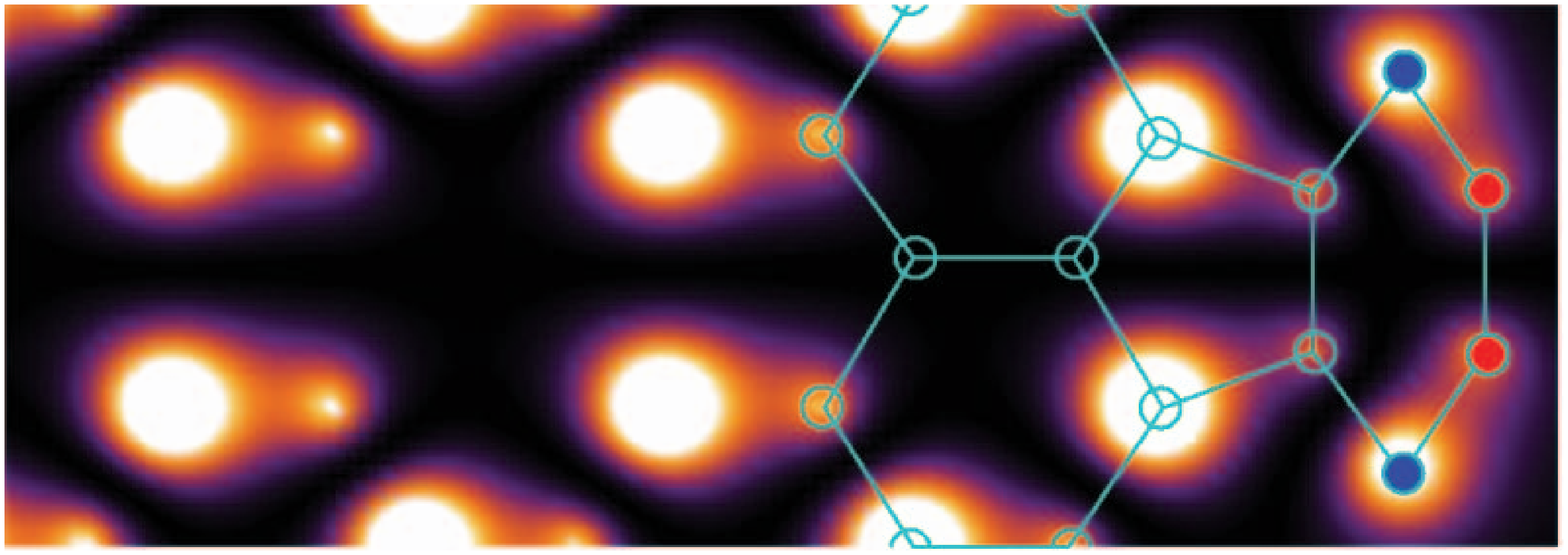}
\caption{ (Color online) ZIGZAG: (a) Half fullerene cap at the end of a $(9,0)$ CNT.
(b) Low energy effective model for the
conducting zigzag nanotube including the half fullerene cap with parameters as
defined in Eq.~(\ref{ZZ_cap_eq}). The last site of the $s$ ($a$) chain corresponds to the $k=0$ ($k=\pi$) mode of the hexagon located at the top of the cap structure.
(c) Local density of states for an $s_{\frac{\pi}{6}}$ state and (d) for an $a_{\frac{\pi}{6}}$ state.}\label{pic_cap_ZZ}
\end{figure}

{}To demonstrate the inclusion of a structure that closes the tube, we focus on the 
case of the $(9,0)$ nanotube depicted in Fig.~\ref{pic_cap_ZZ}(a). 
Such a half fullerene cap can be constructed starting from the perfect CNT in the following way: First, three vacancies are introduced at one end ($l=1$), each one of them producing a pentagon defect as the one discussed above for the vacancy. Then, each one of the $6$ remaining atoms at the end is linked to the atoms that form the hexagon at the top of the cap. Choosing again $\varphi=\frac{\pi}{6}$ it turns out that each one of the aforementioned ``cap vacancies'' (CV) correspond exactly to a term like the one obtained above, in such a way that the part of the cap Hamiltonian corresponding to all three of them is

\bea
H_{CV_{\rm eff}} = 3 H_{V_{\rm eff}}.
\eea
Furthermore, it can be shown that the $s_{\frac{\pi}{6}}$ and $a_{\frac{\pi}{6}}$ channels of the effective model couple only to the $k=0$ and $k=\pi$ modes ($d^{\dagger}_{0, \pi}$) of the hexagon at the top of the cap. The resulting effective Hamiltonian for the cap of the $(9,0)$ nanotube becomes
\bea
\label{ZZ_cap_eq}  
H_{CAP} = H_{CV_{\rm eff}} & +&  2t(d^{\dagger}_{\pi}d^{\phantom\dagger}_{\pi} - d^{\dagger}_{0}d^{\phantom\dagger}_{0}) 
\\ \nonumber
& +& \frac{t}{\sqrt{3}} \left(  s^{\dagger}_{\frac{\pi}{6},1}d^{\phantom\dagger}_{0} - {\rm i}~ \sqrt{3} a^{\dagger}_{\frac{\pi}{6},1}d^{\phantom\dagger}_{\pi} + h.c. \right) , 
\eea
where,
\bea
d_{k} = \frac{1}{\sqrt{6}} \sum_{j=1}^6 d_{j} e^{{\rm i} kj}
\eea
and $d_{j}$ are the sites of the hexagon at the top of the cap.
By conveniently choosing $\varphi=\pi/6$ we have again managed to express the Hamiltonian in terms of two completely decoupled modes even when the cap is present. 
The resulting model requires the addition of a couple of sites and a local change in the values of the parameters as it is illustrated in Fig.~\ref{pic_cap_ZZ}(b).
As for the armchair nanotube, we have described here the
procedure only for one specific cap. Nevertheless, the same scheme can
be followed to model caps of tubes of larger radius that will
naturally require a greater number of extra sites in the effective
model.

The LDOS around the cap of the $(9,0)$ nanotube is displayed in Fig.~\ref{pic_cap_ZZ}(c) and (d) for two different energies near the Fermi level. In this case both $s$ and $a$ states have a finite density at the cap. Notice that due to the $\varphi=\pi/6$ phase shift the forms of the patterns are inverted relative to Fig.~\ref{STM_perfect_ZZ}: the ones of $s_{\frac{\pi}{6}}$ ($a_{\frac{\pi}{6}}$) look like the ones of $a_{0}$ ($s_{0}$).
 This can be easily understood from Eq.~(\ref{C_to_s_a}) since
the shift of  $\varphi=\pi/6$ together with a translation of one site around the tube converts the cosine and sine functions into one another. The effect is clear when comparing Fig.~\ref{STM_perfect_ZZ} with Fig.~\ref{pic_cap_ZZ}.

\section{Defects in the Bulk}   \label{surf_def}
\begin{figure}[tp]
\centering
(a) ~~\includegraphics[width=0.31\textwidth]{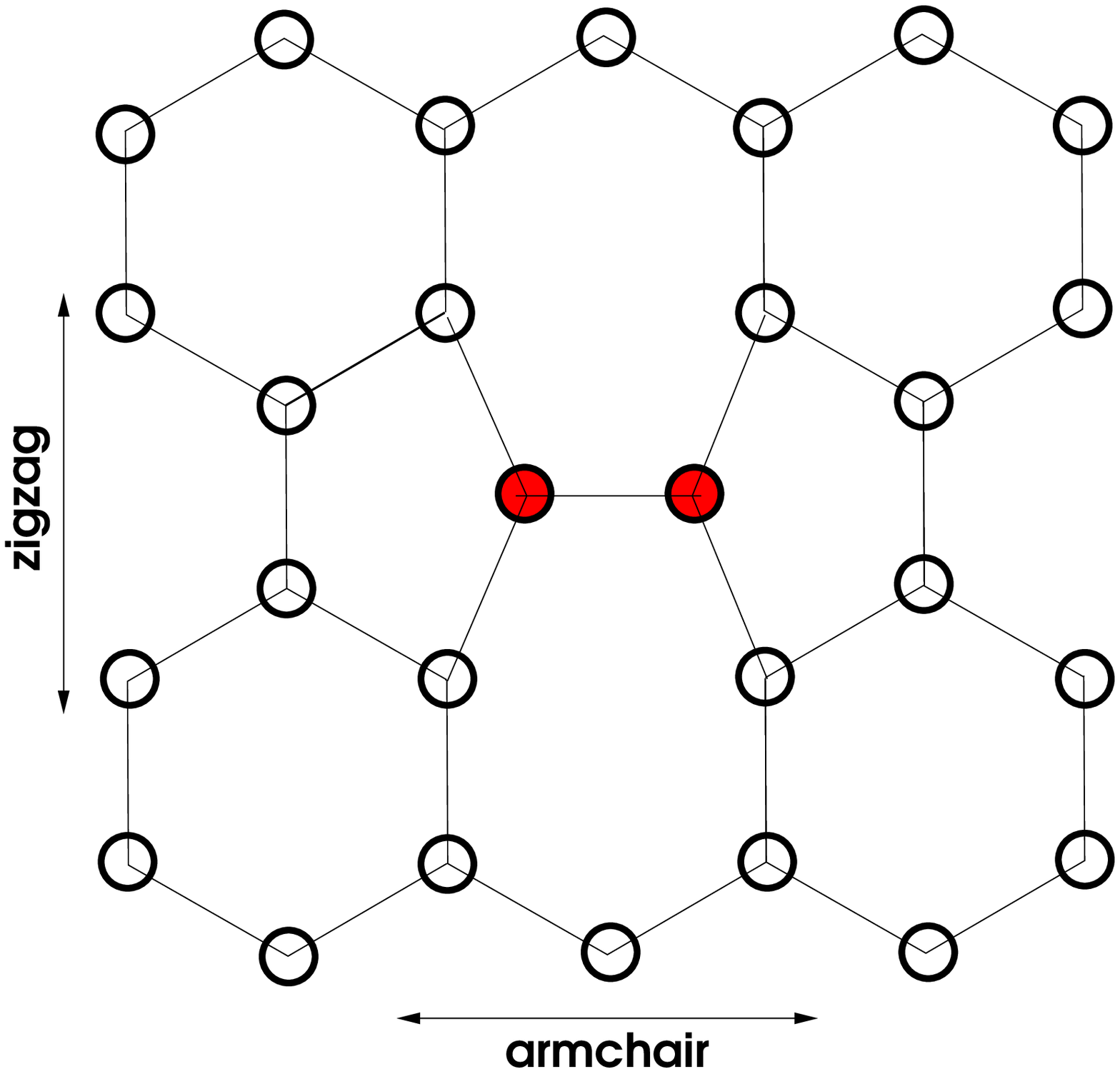} ~~
(b) ~~\includegraphics[width=0.31\textwidth]{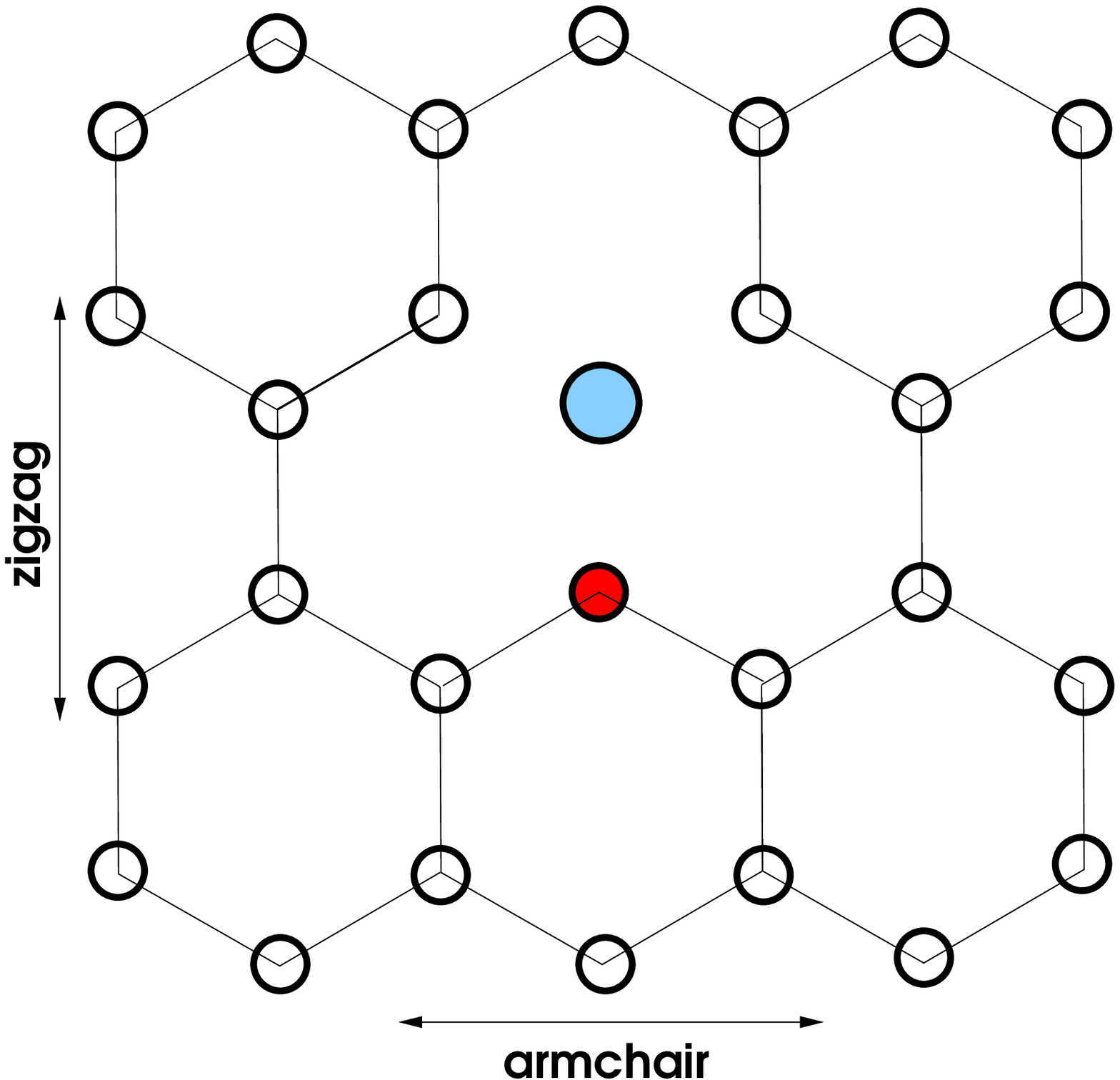} ~~
\caption{(Color online) Two types of defects in the bulk of carbon nanotubes. The arrows indicate the direction of the axis for the two classes of tubes. (a) Stone-Wales defect formed by a pair of pentagon-heptagon deformations. Defective sites are depicted in red and they correspond exactly to the ones added to the effective model. (b) Vacancy defect produced by taking away a single carbon atom, breaking the $\alpha$-$\beta$ sublattice symmetry. 
The big dot (blue) represents the missing atom also in the following figures.}\label{defects}
\end{figure}

The honeycomb structure composing the CNT may present a variety of structural defects which are known to significantly modify its electronic properties.\cite{Choi, Bockrath}
It is possible to model any kind of defect in a straight-forward way using the chain lattice
models introduced in Sec.~\ref{Eff_Mod}.  As an example we will show explicitly how to
 include two prototypical impurity models:
the Stone-Wales and the single vacancy defects as shown in Fig.~\ref{defects}.
We first show how the low energy models look once the defects are included,
and then use them to obtain electron densities for the states closest to the Fermi energy.
As already mentioned in the introduction, relaxation\cite{relax} will in general 
also occur around defects which cannot be implemented in the formalism 
without additional input from 
ab-initio calculations.  Nonetheless, the examples we consider here 
demonstrate the general method also for more complicated cases, and also 
show the most relevant effects that are expected from impurities of this
symmetry class.

\subsection{Stone-Wales defect}
One of the simplest deformations that can be introduced to the perfect graphene lattice is a pair of pentagon-heptagon defects as the one shown in Fig.~\ref{defects}(a). These are usually called Stone-Wales \cite{SW} (SW) or 5-7-7-5 defects and can be thought of as a simple $90^{\circ}$ rotation of a single bond. Stone-Wales defects have low energy and they do not modify the helicity of the tube. They are also known to be responsible for the initiation of the plastic deformation of CNTs as the bond aligns in the direction of the applied strain.\cite{strain}

\subsubsection{Armchair}
\begin{figure}[tp]
\centering
(a) \includegraphics[width=0.44\textwidth]{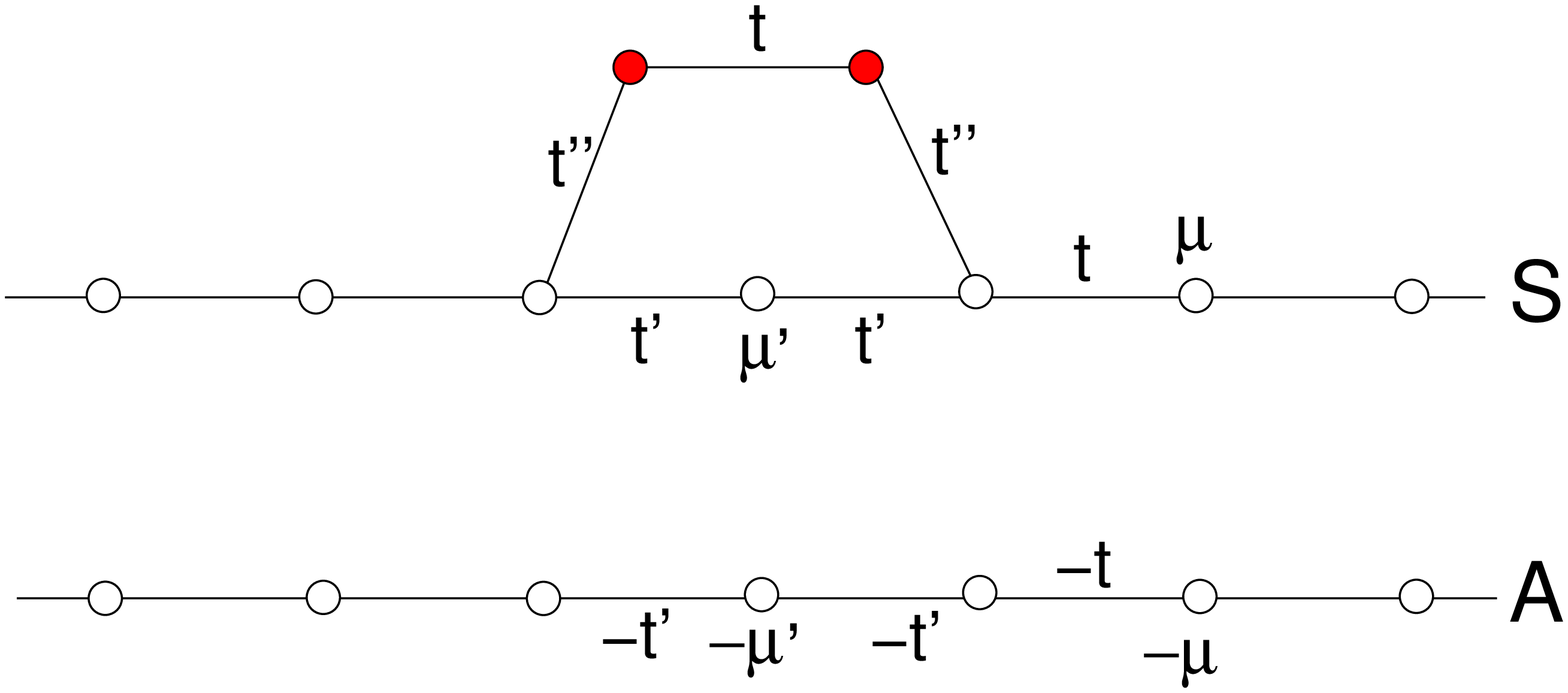}  \\
(b)~ \includegraphics[width=0.44\textwidth]{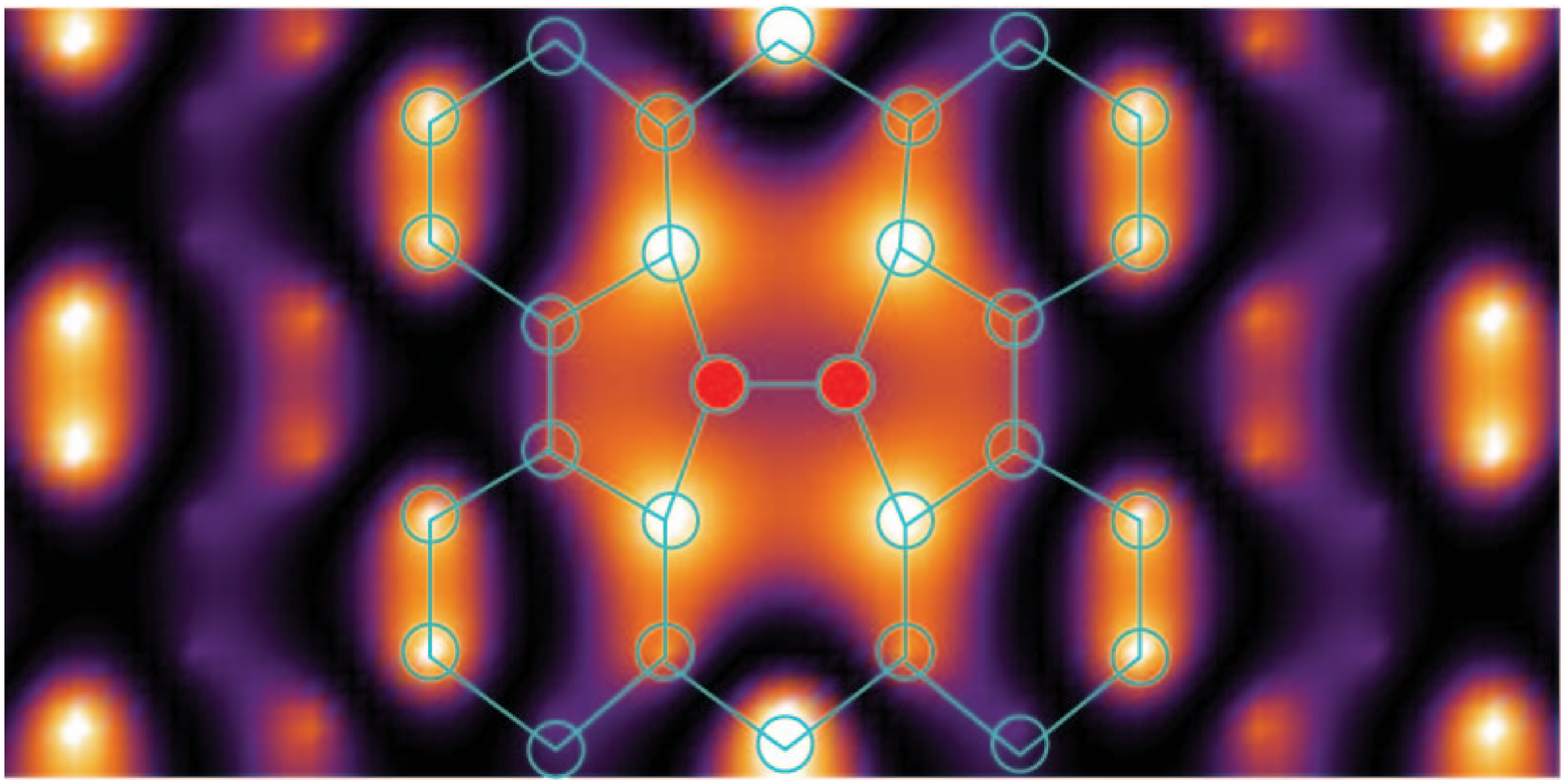}  \\
(c)~ \includegraphics[width=0.44\textwidth]{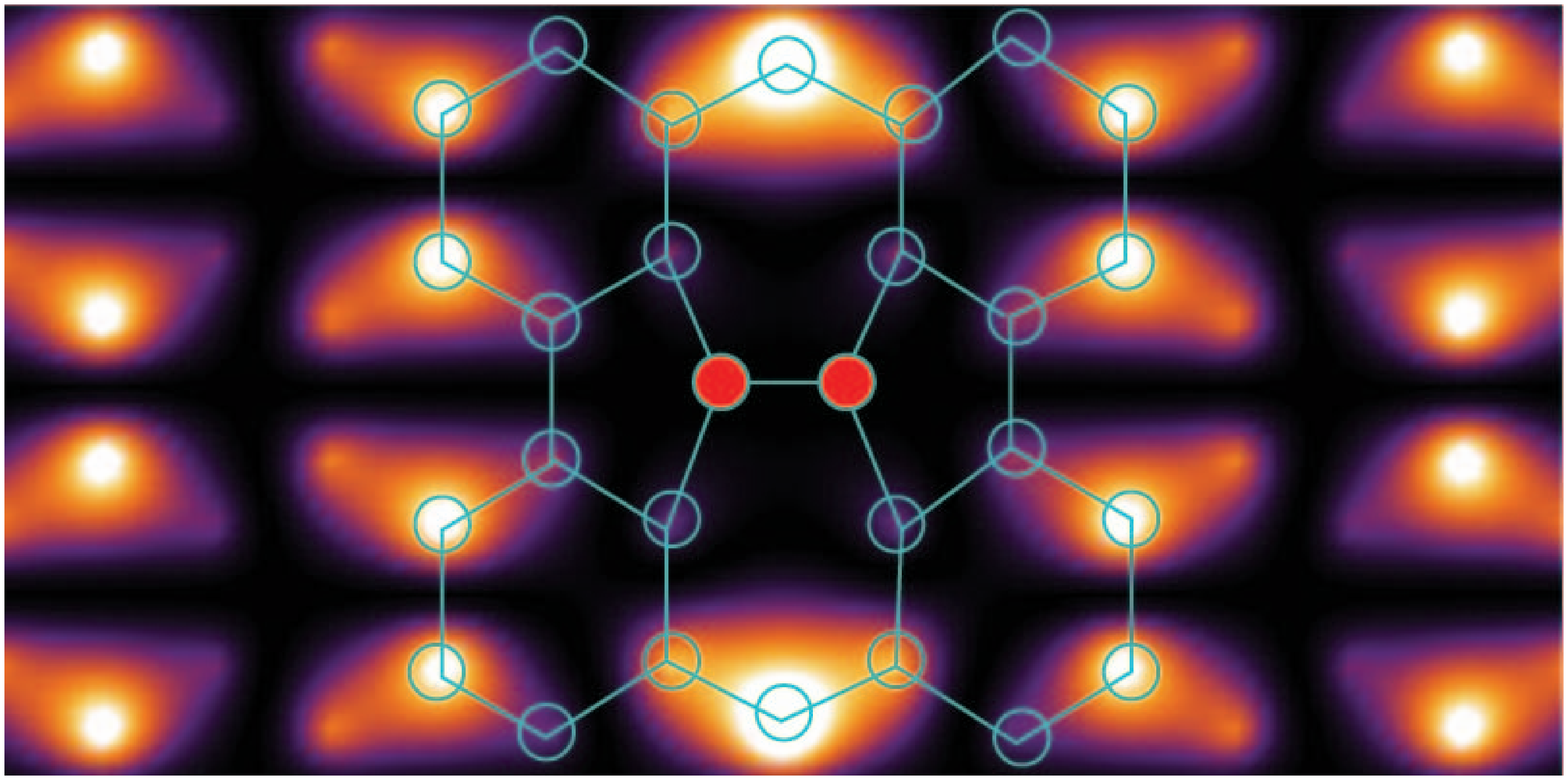}
\caption{(Color online) ARMCHAIR: (a) Effective lattice model including the Stone-Wales defect. The extra sites (in red) correspond to the two carbon atoms of the defective bond. Some hopping amplitudes and chemical potentials are modified around the deformation: $t'=t(N-1)/N,~\mu'=\mu (N-1)/N,~t''=t\sqrt{2/N}$. (b) Electron density around the defect for the $S$ state at the Fermi level. As expected, there is a finite occupation of the defective bond. (c) Same as in (b) but now for the $A$ state. The defective bond is now completely empty.}\label{pic_SW_arm}
\end{figure}

Incorporation of the 5-7-7-5 defect to the effective model of the armchair nanotube is achieved by adding a couple of sites that are connected to the $S$-mode chain 
with the parameters $t'=t(N-1)/N,~\mu'
=\mu (N-1)/N,~t''=t\sqrt{2/N}$ as shown in  
Fig.~\ref{pic_SW_arm}(a). This can be derived in two steps: First, eliminating one bond has the effect of locally weakening the hopping and chemical potential 
of the effective chain sites. 
Then, the  rotated bond is included by adding a couple of extra sites 
corresponding to the atoms of the defective bond. As pointed out above, 
the pentagon defects along the tube are not seen by the antisymmetric mode; 
therefore, the second step does not modify the $A$ chain.

Again, there are interesting effects that can be observed in simulated STM images. Tuning the voltage to an energy corresponding to an eigenstate of the $A$ channel, these sites will be significantly occupied as seen in Fig.~\ref{pic_SW_arm}(b). Otherwise, if the applied voltage matches the energy of the antisymmetric mode, the defective bond will 
appear as completely empty as shown in  Fig.~\ref{pic_SW_arm}(c). 
In both cases, away from the defect the density pattern will look 
essentially identical to the one of the perfect tube as shown in 
Fig.~\ref{STM_perfect_ARM}.

\begin{figure}[tp]
\centering
(a) \includegraphics[width=0.44\textwidth]{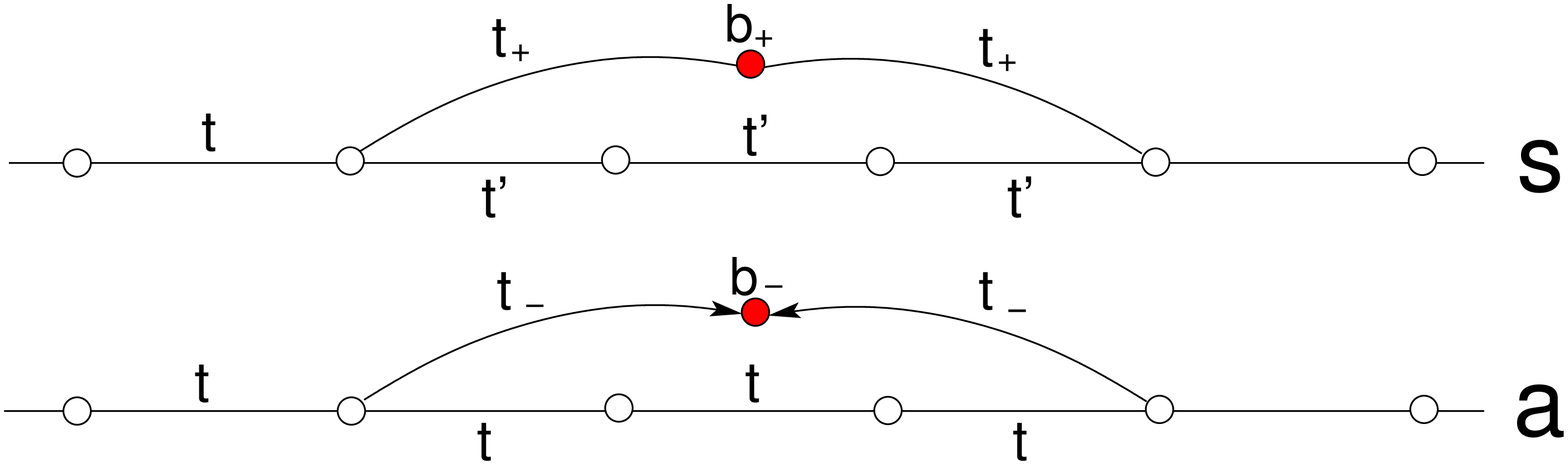}  \\
(b)~ \includegraphics[width=0.44\textwidth]{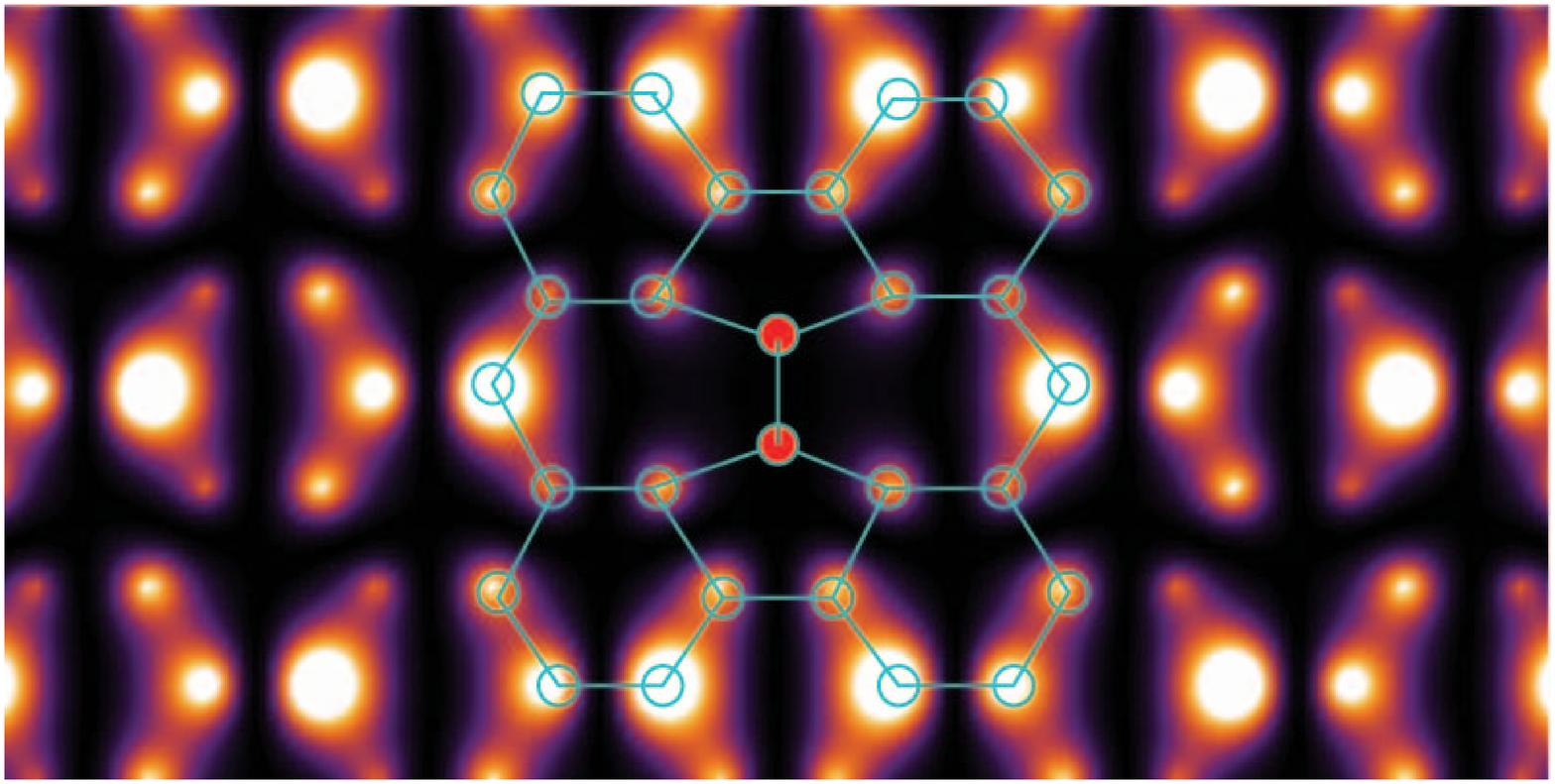}  \\
(c)~ \includegraphics[width=0.44\textwidth]{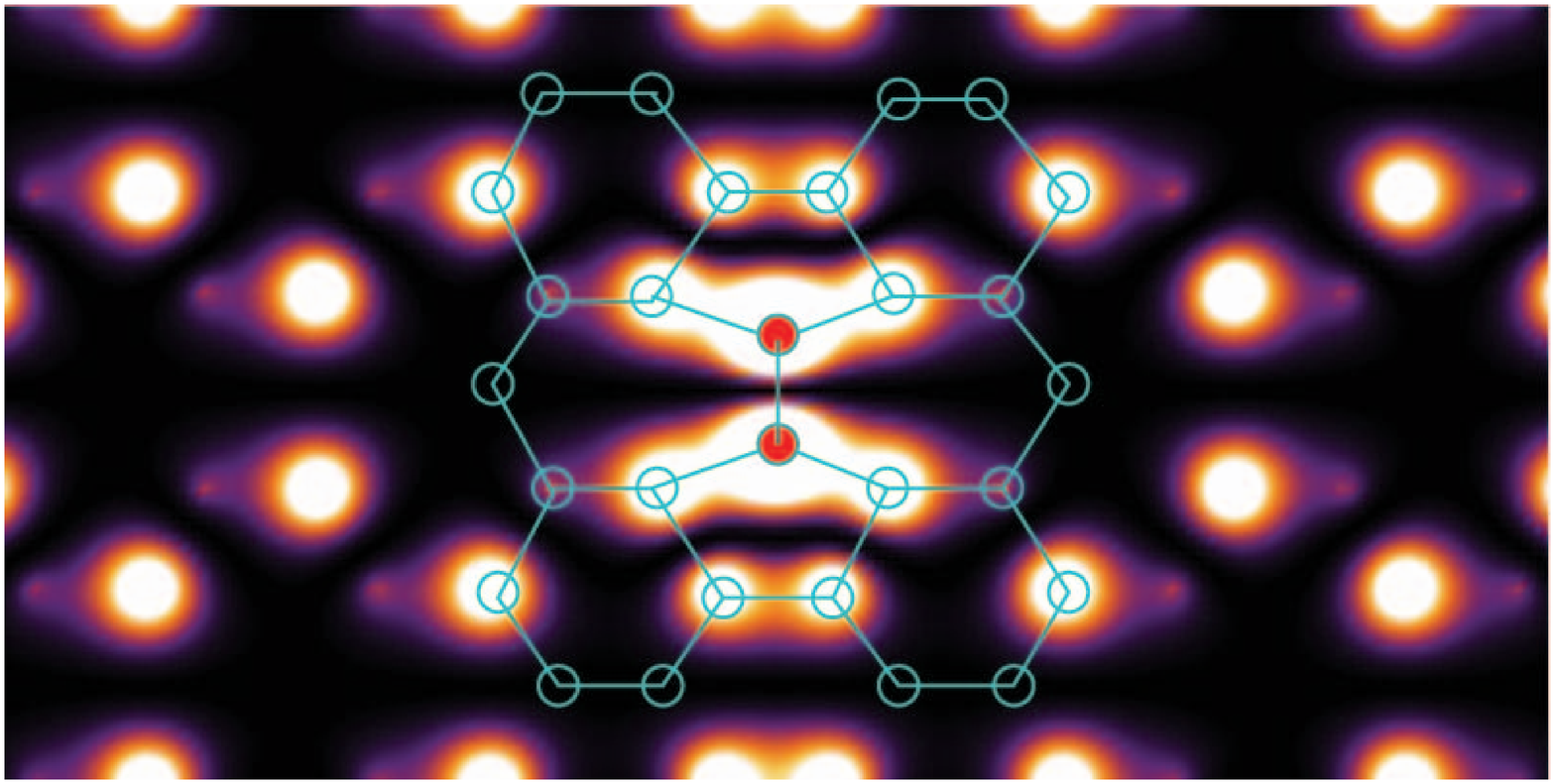}
\caption{ (Color online) ZIGZAG: (a) Effective lattice model for the zigzag tube including the Stone-Wales defect. Additional sites correspond to the symmetric ($b_{+}$) and antisymmetric ($b_{-}$) modes of the defective bond. The values for the locally modified parameters are
 $t'=t(N-1)/N$, $t_{+}=t/\sqrt{N}$ and $t_{-}=\pm {\rm i} \sqrt{3} t_{+}$. The hopping direction in which $t_{-}$ has a positive sign is indicated by arrows to $b_{-}$. (b) Electron density around the defect for a $s_{\frac{\pi}{6}}$ state close to the Fermi level.  (c) Same as in (b) but now for a $a_{\frac{\pi}{6}}$ state.}\label{pic_SW_ZZ}
\end{figure}

\subsubsection{Zigzag}
The situation for the zigzag tube is rather different. First we choose $\varphi=\pi/6$ for the modes $s_{\varphi}$ and $a_{\varphi}$ in the 
transformation (\ref{C_to_s_a}), since it yields the simplest possible low energy model. In such a basis the elimination of a bond translates into  a weakening of a few hopping elements of the $s$ chain but leaves $a$ intact. Furthermore, the subsequent addition of the defective bond involves connecting the $s$ ($a$) channel to the $b_{+}$ ($b_{-}$) mode of the added bond, where $b_{\pm}$ is defined by Eq.~(\ref{bond_trafo}). The resulting form of the effective model with the parameters $t'
=t(N-1)/N$, $t_{+}=t/\sqrt{N}$ and $t_{-}=\pm {\rm i} \sqrt{3} t_{+}$
is shown in Fig.~\ref{pic_SW_ZZ}(a).

Patterns obtained from an STM experiment around the imperfection for energies close to the Fermi level are expected to look like Fig.~\ref{pic_SW_ZZ}(b) and (c). Depending on the symmetry of the state and the position of the defect, the Stone-Wales bond can be observed to accumulate a large amount of density as in Fig.~\ref{pic_SW_ZZ}(b) or in other cases become completely empty as 
in Fig.~\ref{pic_SW_ZZ}(c). Unlike for the armchair case, in the zigzag CNT 
the lack of density on the defective bond is not an 
exclusive property of a state with a certain symmetry.

\subsection{Vacancy defect}

Irradiation of a carbon nanotube by Ar$^+$ ions may result in the presence of metastable but long-lived single vacancy defects on its surface \cite{Krash_vac} as illustrated in Fig.~\ref{defects}(b). The absence of an atom in the carbon network obviously does not change the overall chiral vector of the tube, but it will still have some pronounced consequences on its electronic properties. For example, recent theoretical\cite{Biel_And_loc} and experimental\cite{Kong_transp, Gomez_And_loc} works have shown that nanotubes enter the strong Anderson localization regime when the concentration of defects is large enough.

\begin{figure}[tp]
\centering
(a) \includegraphics[width=0.45\textwidth]{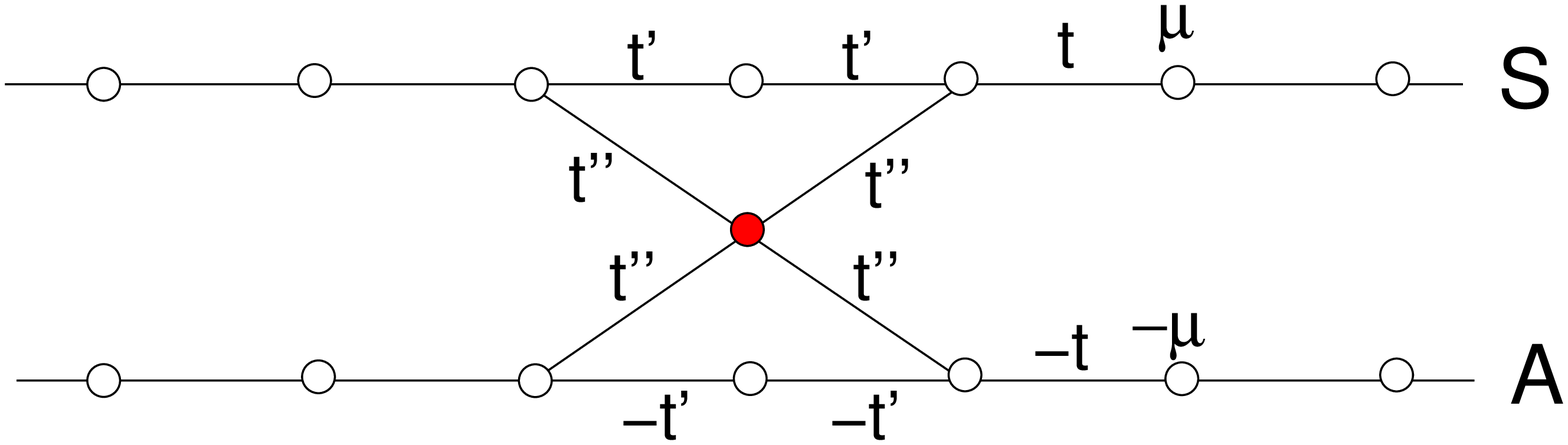}  \\
(b)~ \includegraphics[width=0.41\textwidth]{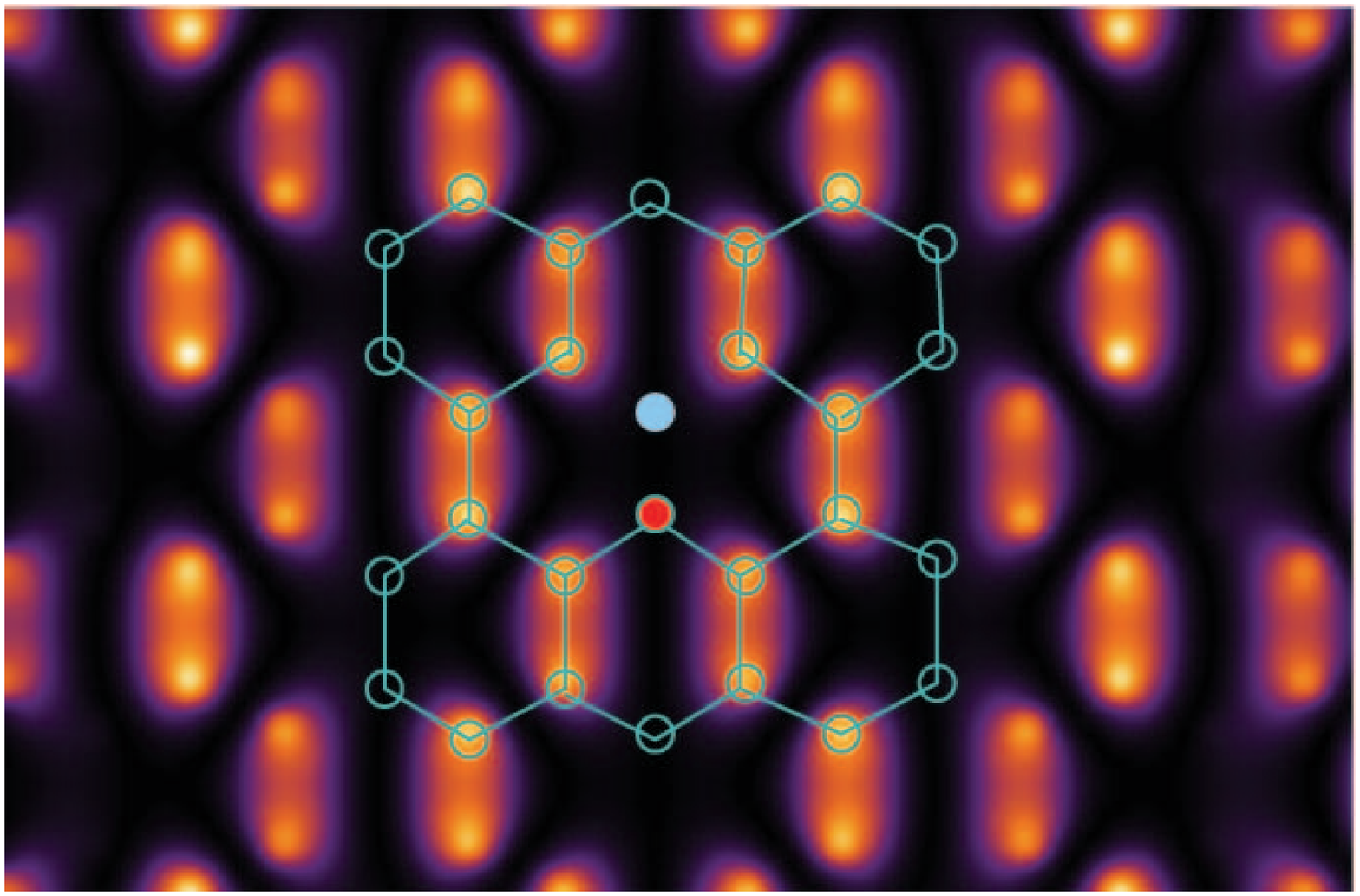}  \\
(c)~ \includegraphics[width=0.41\textwidth]{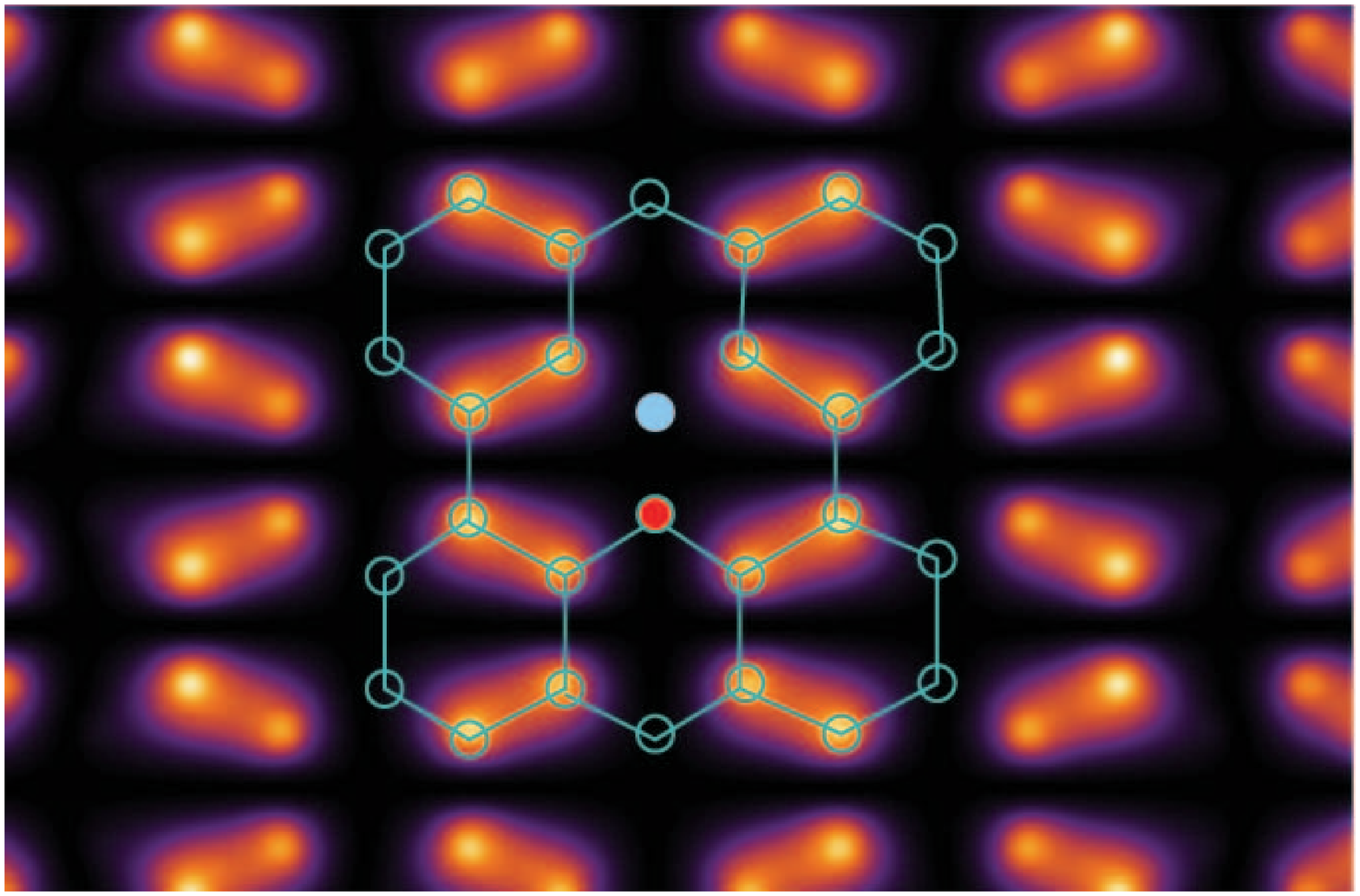} \\
(d)~ \includegraphics[width=0.41\textwidth]{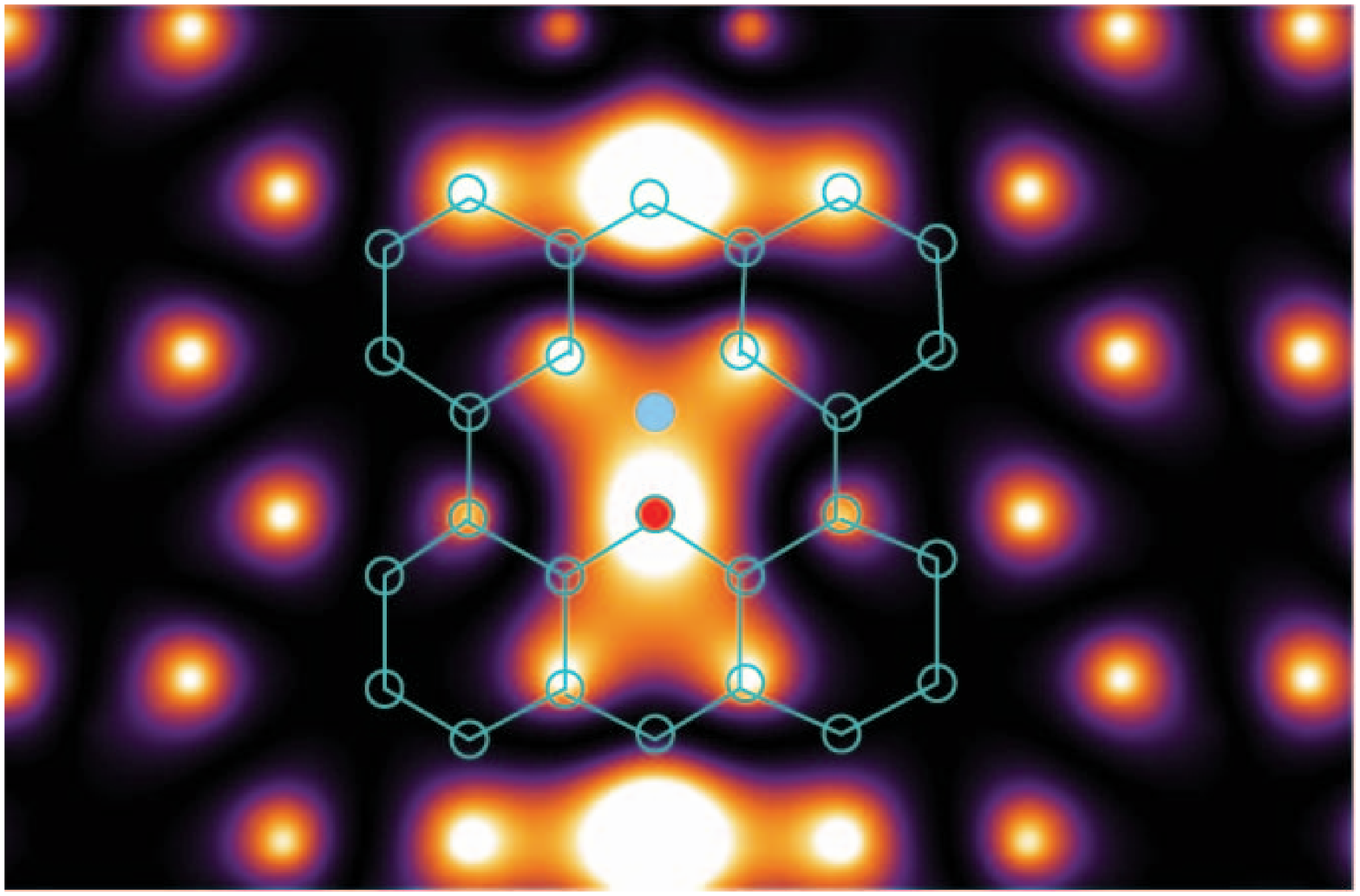}
\caption{ (Color online) ARMCHAIR: (a) The vacancy is included into the low energy effective lattice model. An extra site represents the nearest neighbor of the vacancy in the direction perpendicular to the tube's axis. Parameters are modified around the imperfection: $t'=t(N-1)/N,~\mu'=\mu (N-1)/N,~t''=t/\sqrt{2N}$. (b) Electron density around the defect for an eigenstate which looks basically exactly the same as one corresponding to the pure symmetric channel ($S$-like). (c) Same as in (b) but now the density looks as the one of the $A$ mode ($A$-like). (d) There is a third group of eigenstates that combine both channels in such a way that the density is accumulated in one of the sublattices; this is a manifestation of the 
broken sublattice symmetry by the vacancy defect.}\label{pic_vac_arm}
\end{figure}

\subsubsection{Armchair}  \label{surf_def_SW_arm}
To include the vacancy in the effective chain model a single {\it extra} 
site must be added, which corresponds to the remaining atom of the defective bond 
with coupling parameters $t'=t(N-1)/N,
~\mu'=\mu (N-1)/N,~t''=t/\sqrt{2N}$ as shown in
Fig.~\ref{pic_vac_arm}(a). It is interesting to note that this time the $S$ and $A$ chains become connected as a consequence of the introduction of the defect. Therefore, eigenstates will not necessarily correspond to the electron occupying a single mode, but they will in general be a mixture of both. It turns out that there are now three types of possible 
states: $S$-like, $A$-like and mixed ($M$). The first two correspond to states in which the electron occupies primarily only one of the original channels as illustrated in 
Fig.~\ref{pic_vac_arm}(b) and (c). Mixed states are also possible, where the eigenstate is a linear combination of the $S$ and $A$ modes. It is found that the $M$ states completely break the sub-lattice symmetry of the hexagonal lattice, resulting in an accumulation of the electron density on the sublattice opposite to the one of the vacancy as shown in 
Fig.~\ref{pic_vac_arm}(d).
It appears that the effective extra site enables a resonance at certain energies.  
A related effect of a strongly enhanced spectral weight around vacancies 
at certain energies has been predicted for graphene using a linearized band 
structure.\cite{graphene}

For quantitative estimates 
more realistic microscopic models must also take relaxation 
effects into account which are especially important for vacancy defects.  
This will generally modify the couplings and STM images in a finite range, but 
the resulting eigenstates can be classified into one of the three types (A), (S), or (M)
with the characteristic general properties as shown in Fig.~\ref{pic_vac_arm}.

\subsubsection{Zig-zag}
\begin{figure}[tp]
\centering
(a) \includegraphics[width=0.44\textwidth]{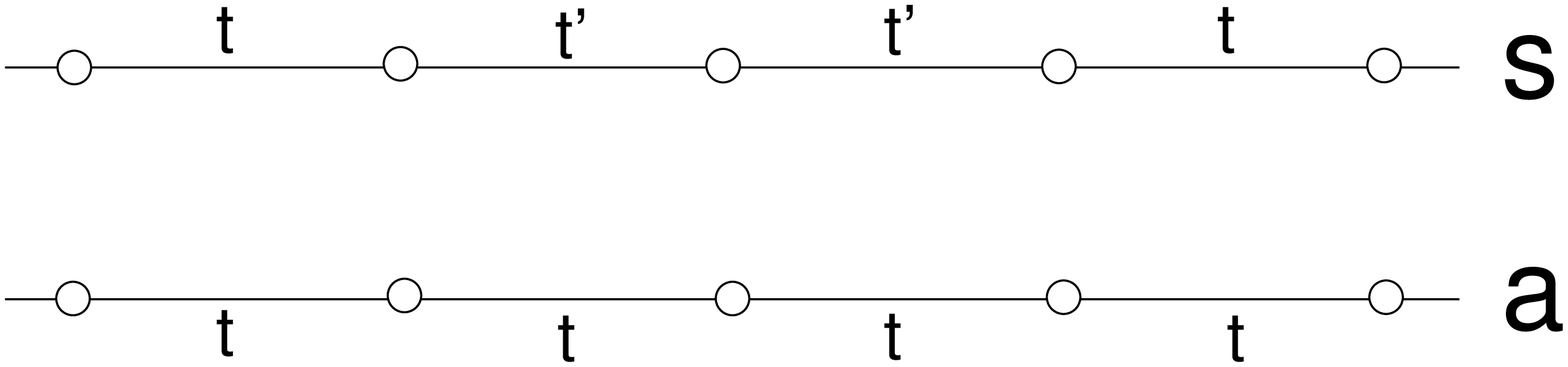}  \\
(b)~ \includegraphics[width=0.44\textwidth]{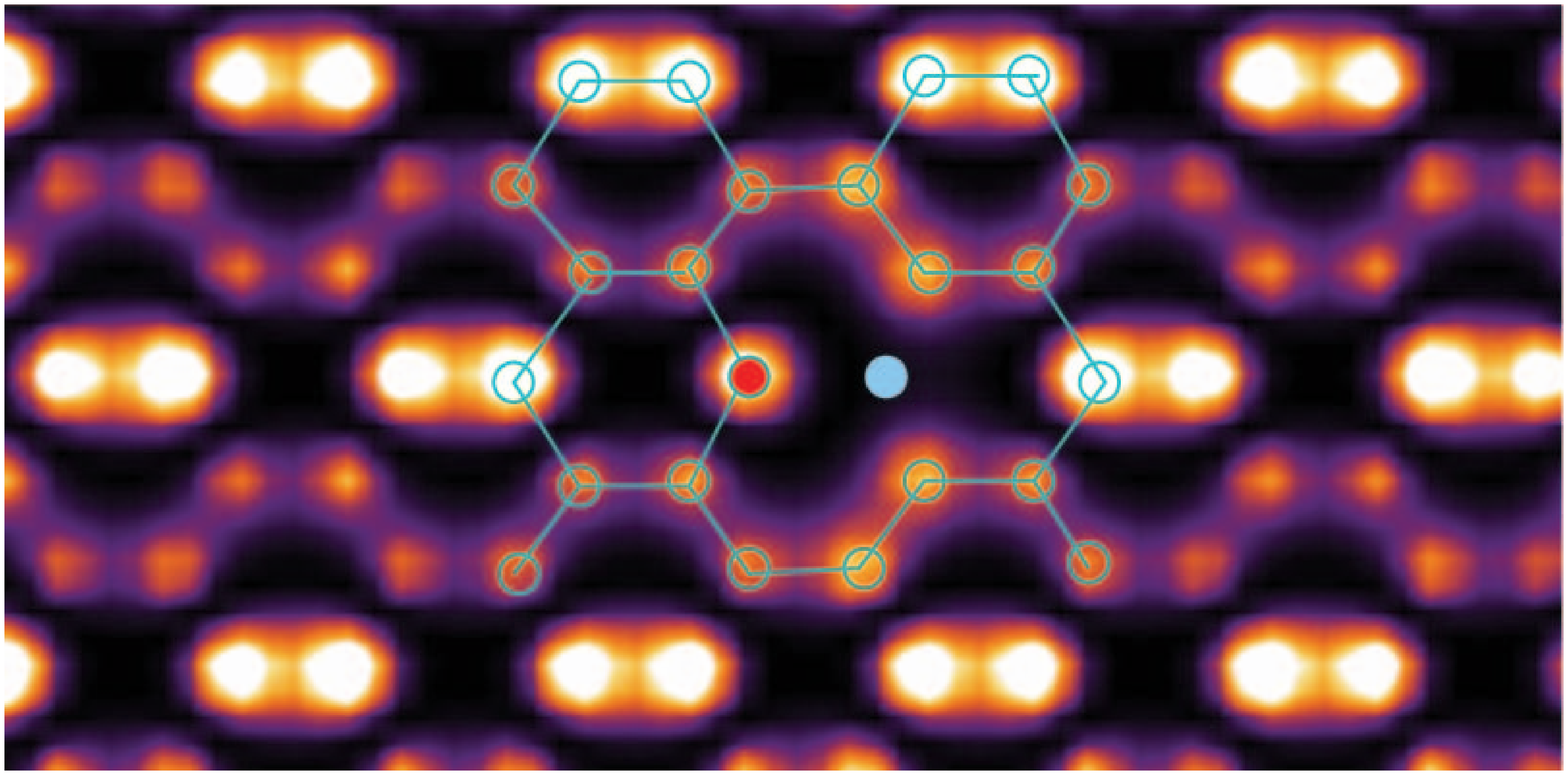}  \\
(c)~ \includegraphics[width=0.44\textwidth]{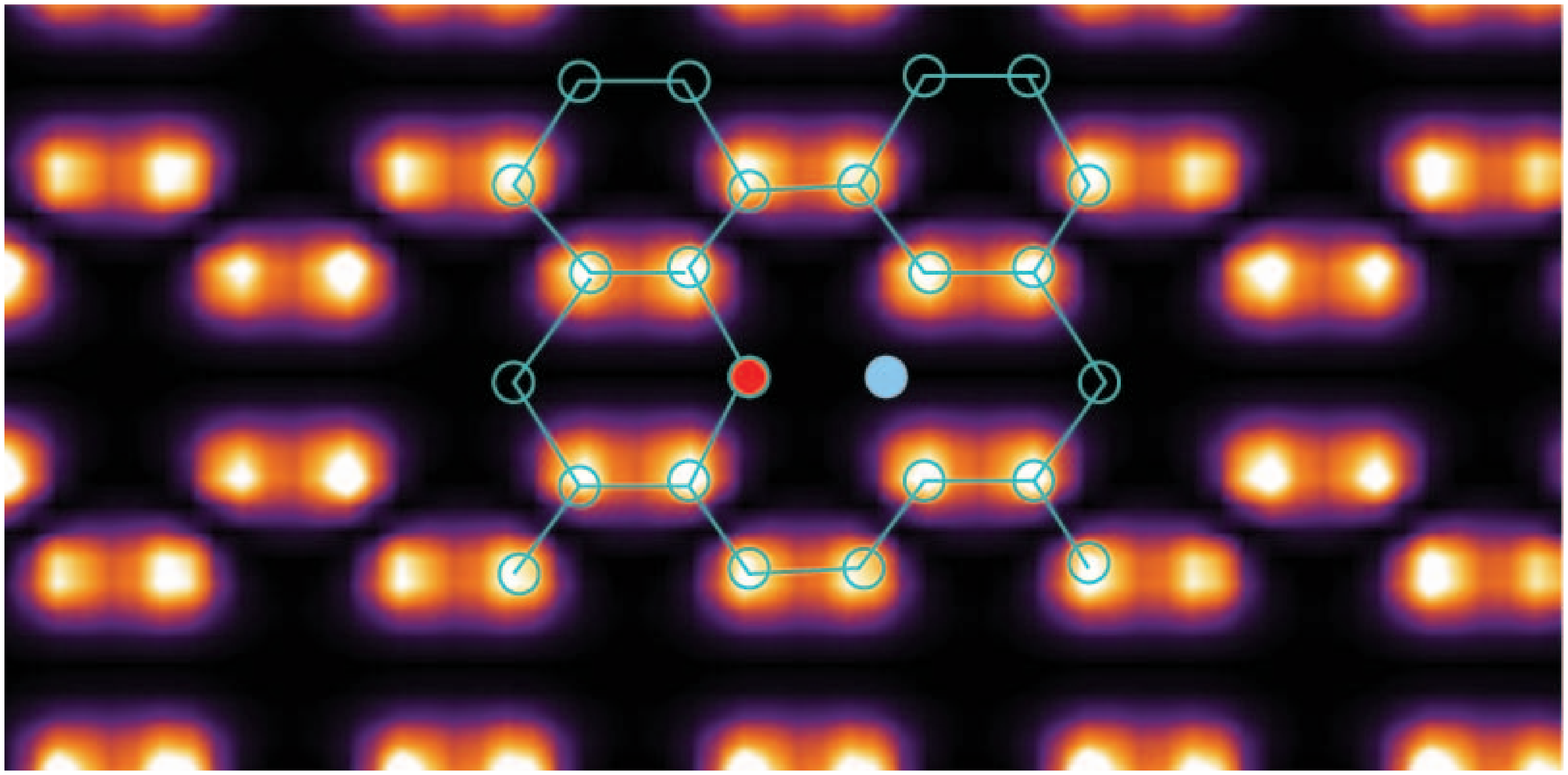}
\caption{ (Color online) ZIGZAG: (a) Chain lattice model including the vacancy defect. There is a slight
weakening of the hopping along the $s_{\frac{\pi}{6}}$ channel from $t$ to $t'=t(N-1)/N$. (b) Local density of states around the imperfection for a level corresponding to the $s_{\frac{\pi}{6}}$ chain. (c) Density pattern around the vacancy for an $a_{\frac{\pi}{6}}$ state.}\label{pic_SW_zig}
\end{figure}

For the case of the zigzag nanotube, the vacancy  
does not require an additional site in the effective model since it represents only a slight modification 
on one effective chain lattice site. In this case, the choice $\varphi=\pi/6$ again yields 
the simplest effective model.  In fact only the $s_{\frac{\pi}{6}}$ mode is modified 
by the presence of a vacancy which causes 
a slight weakening of local hopping terms $t'=t(N-1)/N$ as shown in 
Fig.~\ref{pic_SW_zig}(a). 
We note that the same impurity terms were also present when we considered the Stone-Wales 
defect, which is not surprising given that it 
is nothing more than two consecutive vacancies plus the attachment of the defective bond.

As seen in Fig.~\ref{pic_SW_zig}(b) and (c), the broken sublattice symmetry is not 
manifest in the density patterns. Nevertheless, the presence of the vacancy has the 
effect of pinning the phase of the wavefunction around the nanotube. 
Remarkably, the antisymmetric channel is not at all affected by the vacancy, except of course
for the missing spectral weight at the vacancy.  As discussed before, the choice of 
$\varphi=\pi/6$ makes the antisymmetric state appear like the $\varphi=0$ symmetric state
in Fig.~\ref{STM_perfect_ZZ} and vice versa.

As mentioned above relaxation is important also in the zig-zag case.  However, since
the parity symmetry of the system is not broken by a vacancy in the zig-zag tube
 the channels are bound to remain
decoupled and the general classification in Fig.~\ref{STM_perfect_ZZ} still holds.

\section{Summary and Outlook}
In summary, we have shown how defects and capping structures can be added to 
effective two chain models of conducting SWCNTs. We have demonstrated the method by incorporating half fullerene caps and Stone-Wales and vacancy defects to both the armchair and conducting zigzag nanotubes. Using the resulting chain lattice models of the conduction bands
we obtained the LDOS around the imperfections for states close to the Fermi level.

Besides their simplicity the most remarkable feature of the effective models 
is the fact that the two chains are completely decoupled and also remain decoupled even
in the presence of most defects and additional structures at the end considered here.
Especially in the zigzag case, the defects in fact dictate the proper choice of the 
basis, which then represent the two independent channels.
Surprisingly, a single vacancy leaves all states in 
the antisymmetric channel of metallic zigzag CNTs completely unchanged. 
Therefore, transport in that channel should be possible without scattering,
which may have
strong consequences for the resulting conductivity.  However, 
this reduction of scattering is only 
possible if vacancies are spaced so that they do not affect each other and if there are
no other strong perturbations which would pin the phase $\varphi$ around the tube.
Interactions will introduce a coupling between the independent chains as discussed 
in the appendix, but of course even in that case it is useful to 
start in the basis of decoupled chains.

At this point it should be emphasized that the methods presented 
above are not restricted to the examples explicitly shown here,
but the transformations in Sec.~\ref{Eff_Mod}
give a quite general framework which allows for the treatment of a
wide range of possible imperfections. For example, the procedure illustrated in the present paper can be extended to tackle problems such as junctions of (conducting) CNTs of different chirality, presence of magnetic impurities, or transport related problems, just to mention a few interesting prospects.

{}From the point of view of applications, it is clear that this
approach can be useful to assess 
the effect that a particular deformation or impurity has on the electronic 
properties of nanotubes in a very simple and precise manner. 
It is thus a promising tool 
to be employed for the design of CNT based microelectronic devices.

On the other hand, from a more fundamental perspective, since the resulting models are one-dimensional, it will be certainly interesting 
to take into account strong correlation effects in 
future works by using the techniques available for such 
systems ({\it i.e.}~bosonization or DMRG).

\begin{acknowledgments}
We are thankful for discussions with Michael Bortz, Imke Schneider, and Stefan S\"offing.
Financial support was granted by the Deutsche Forschungsgemeinschaft via the 
Transregio 49 "Condensed Matter Systems with Variable Many-Body Interactions".
\end{acknowledgments}

\appendix
\section{Interactions}
\label{Interactions}
In the main text we have neglected interactions between electrons and worked exclusively with the resulting free electron models. We will now show that including interactions into the effective Hamiltonian can be done in a straight-forward way. Although in principle we could consider any electron-electron potential, for definiteness we will treat only on-site density-density interactions ($U$). To illustrate how a defect modifies the interacting term, we will explicitly show the case of the Stone-Wales deformation.

\subsection{Armchair}
For the armchair carbon nanotube an on-site interaction is given by
\bea
H^{\rm int} = U \sum_{\vec{r}} \left(\alpha^{\dagger}_{\vec{r}, \uparrow}
\alpha_{\vec{r}, \uparrow} \alpha^{\dagger}_{\vec{r}, \downarrow}\alpha_{\vec{r}, \downarrow} + \beta^{\dagger}_{\vec{r}, \uparrow}\beta_{\vec{r}, \uparrow} \beta^{\dagger}_{\vec{r}, \downarrow}\beta_{\vec{r}, \downarrow} \right) .
\eea
Using Eq.~(\ref{FT_armchair}) the conduction band part becomes,
\bea
H^{\rm int}_{\rm eff} = \frac{U}{N} \sum_{y} \left(\alpha^{\dagger}_{y,\uparrow}\alpha_{y,\uparrow} \alpha^{\dagger}_{y,\downarrow}\alpha_{y,\downarrow} + \beta^{\dagger}_{y,\uparrow}\beta_{y,\uparrow} \beta^{\dagger}_{y,\downarrow}\beta_{y,\downarrow} \right).
\eea
In terms of the $S$ and $A$ modes this term looks slightly more complicated,
\bea\label{armint}
H^{\rm int}_{\rm eff} = \frac{U}{4N} \sum_{y}& &\!\!\!\!\!\!\!\!\!\!  \left(n_{SS}^{\uparrow}(y)n_{SS}^{\downarrow}(y) + n_{AA}^{\uparrow}(y)n_{AA}^{\downarrow}(y) \right.
\\  \nonumber
& +&  n_{SA}^{\uparrow}(y)n_{SA}^{\downarrow}(y) + n_{AS}^{\uparrow}(y)n_{AS}^{\downarrow}(y) 
 \\ \nonumber 
& +&  n_{SS}^{\uparrow}(y)n_{AA}^{\downarrow}(y) + n_{AA}^{\uparrow}(y)n_{SS}^{\downarrow}(y) 
\\ \nonumber
& +& \left. n_{SA}^{\uparrow}(y)n_{AS}^{\downarrow}(y) + n_{AS}^{\uparrow}(y)n_{SA}^{\downarrow}(y)\right), 
\eea
where,
\bea
n_{P Q}^{\sigma} = P^{\dagger}_{\sigma} Q_{\sigma}
\eea
As we can see, even though rewriting the Hamiltonian in terms of the $S$ and $A$ modes simplifies its non interacting part, the interaction terms become more convoluted. It is interesting to note that the effective on-site interaction becomes inversely proportional to the number of bonds around the tube ($U_{\rm eff} = U/4N$). Such a dependence is not surprising: the occupation of a site in one of the effective low energy channels represents the presence of a single electron distributed around the tube's $N$ bonds.

To include the Stone-Wales defect we first note that a (normal) bond around the nanotube is missing and there is thus a slight decrease of the on-site interaction at the corresponding site,
\bea
U_{\rm eff}(y_d) = \frac{(N-1)}{N} U_{\rm eff}.
\eea
Furthermore, we have to add the interaction terms that act on the added (defective) bond,
\bea
H_{b}^{\rm int} = U (n_{b_1}^{\uparrow} n_{b_1}^{\downarrow} + n_{b_2}^{\uparrow} n_{b_2}^{\downarrow}).  \label{bond_int}
\eea
Remarkably, while the interaction in the bulk of the tube is reduced by a factor of $1/N$
in Eq.~(\ref{armint}),
the effective interaction on the Stone-Wales bond is not rescaled and therefore
is bound to play a much stronger role.

\subsection{Zigzag}
The interaction Hamiltonian for the zigzag CNT is,
\bea
H^{\rm int} = U \sum_{j,l} c^{\dagger}_{j,l \uparrow}c_{j,l \uparrow} c^{\dagger}_{j,l \downarrow}c_{j,l \downarrow} .  \label{ZZ_int}
\eea
Using the two band approximation and transformation in Eq.~(\ref{C_reverse})
the effective interaction term becomes,
\bea
H^{\rm int}_{\rm eff}  =  \frac{U}{N} \sum_{l} &&\!\!\!\!\!\!\!\!\!\! \left(n_{++}^{\uparrow}(l)n_{++}^{\downarrow}(l) + n_{--}^{\uparrow}(l)n_{--}^{\downarrow}(l) \right. \\  \nonumber
& +&  n_{++}^{\uparrow}(l)n_{--}^{\downarrow}(l) + n_{--}^{\uparrow}(l)n_{++}^{\downarrow}(l) 
\\ \nonumber
& +&  \left.  n_{+-}^{\uparrow}(l)n_{-+}^{\downarrow}(l) + n_{-+}^{\uparrow}(l)n_{+-}^{\downarrow}(l) \right),
\eea
where,
\bea
n_{\mu \nu}^{\sigma} = C_{\mu, \sigma}^{\dagger} C_{\nu, \sigma}.
\eea
We find that the effective interaction is again inversely proportional to the tube's circumference ($U_{\rm eff}=U/N$). Note also that the initially independent chains of (\ref{ZZ_eff_H}) become correlated by the scattering terms that involve both $+$ and $-$ modes.

The first effect that the Stone-Wales defect will have on this model is a weakening of the interaction strength ($U_{\rm eff}$) similar to the one derived for the armchair tube, but now it affects two contiguous sites, $l_d$ and $l_d+1$ (see Fig.~\ref{pic_SW_ZZ}). They will both have,
\bea
\tilde U_{\rm eff} = \frac{(N-1)}{N} U_{\rm eff}.
\eea
Furthermore, a couple of terms of order $O(N^{-2})$ arise, which contain local interaction vertices of a kind that was not present for the pure system,
\bea
\tilde{H}^{\rm int}_l  =  -\frac{U}{N^2} &&\!\!\!\!\!\!\!\!\!\!\left[ n_{+-}^{\uparrow}n_{+-}^{\downarrow}(l) + n_{-+}^{\uparrow}n_{-+}^{\downarrow}(l) \right. \\
& -& (n_{++}^{\uparrow} + n_{--}^{\uparrow})(n_{+-}^{\downarrow} + n_{-+}^{\downarrow})(l)   
\nonumber  \\
& -& \left. (n_{++}^{\downarrow} + n_{--}^{\downarrow})(n_{+-}^{\uparrow} + n_{-+}^{\uparrow})(l)  \right],~~~~~ \nonumber
\eea
one for each of the defective sites mentioned before ($l=l_d,l_d+1$).

The last and most trivial part of the interaction term coming from the impurity corresponds to the on-site terms of the defective bond. Such a term will be exactly the same as the one given for the armchair carbon nanotube in (\ref{bond_int}), again with a not rescaled 
stronger interaction strength.


\begin{thebibliography}{10}
\bibitem{IijimaSW}
S.~Iijima and T.~Ichihashi.
 {Nature (London)} {\bf 363}, 603, (1993).


\bibitem{Bethune}
D.~S. Bethune, C.~H. Kiang, M.~S. de~Vries, G.~Gorman, R.~Savoy, J.~Vazquez,
  and R.~Beyers.
 {Nature (London)} {\bf 363}, 605, (1993).


\bibitem{book}  For a review on the electronic properties of SWCNTs see
J.-C.~Charlier, X.~Blase, S. Roche, Rev. Mod. Phys. {\bf 79}, 677 (2007).


\bibitem{Kane}
C.L.~Kane and E.J.~Mele.
 {Phys. Rev. Lett.} {\bf 78}, 1932, (1997).


\bibitem{gap}
A.~Kleiner and S.~Eggert, Phys. Rev. B {\bf 63}, 073408 (2001).


\bibitem{Rubio}
A.~Rubio, D.~Sanchez-Portal, E.~Artacho, P.~Ordejon, and J.~M. Soler.
 {Phys. Rev. Lett.} {\bf 82}, 3520, (1999).


\bibitem{Rochefort}
A.~Rochefort, D.~R. Salahub, and P.~Avouris.
 {J. Phys. Chem. B} {\bf 103}, 641, (1999).


\bibitem{Jiang}
J.~Jiang, J.~Dong, and D.Y.~Xing.
 {Phys. Rev. B} {\bf 65}, 245418, (2002).


\bibitem{Venema}
L.~C. Venema, J.~W.~G. Wildoer, J.~W. Janssen, S.~J. Tans, H.~L. J.~Temminck
  Tuinstra, L.~P. Kouwenhoven, and C.~Dekker.
 {Science} {\bf 283}, 52, (1999).


\bibitem{Lemay}
S.~G. Lemay, J.~W. Jansen, M.~van~den Hout, M.~Mooij, M.~J. Bronikowski, P.~A.
  Willis, R.~E. Smalley, L.~P. Kowenhoven, and C.~Dekker.
 {Nature (London)} {\bf 412}, 617, (2001).


\bibitem{anfuso} F.~Anfuso and S.~Eggert, Phys.~Rev.~B \textbf{68}, 241301(R) (2003)


\bibitem{schneider}
I.~Schneider, A.~Struck, M.~Bortz, and S.~Eggert,
 Phys.~Rev.~Lett. {\bf 101}, 206401 (2008)


\bibitem{lee}
J.~Lee, S.~Eggert, H.~Kim, S.-J.~Kahng, H.~Shinohara, Y.~Kuk,
Phys.~Rev.~Lett.~\textbf{93}, 166403 (2004).


\bibitem{stm}
S.~Eggert, Phys.~Rev.~Lett. {\bf 84}, 4413 (2000).


\bibitem{Fan}
Y.~F. Fan, B.~R. Goldsmith, and P.~G. Collins.
 {Nature Materials} {\bf 4}, 906, (2005).


\bibitem{Charlier}
J.-C. Charlier, T.W.~Ebbesen, and Ph.~Lambin.
 {Phys. Rev. B} {\bf 53}, 11108, (1996).


\bibitem{Chico}
L.~Chico, L.~X. Benedict, S.~G. Louie, and M.~L. Cohen.
 {Phys. Rev. B} {\bf 54}, 2600, (1996).


\bibitem{Choi}
H.~J. Choi, J.~Ihm, S.~G. Louie, and M.~L. Cohen.
 {Phys. Rev. Lett.} {\bf 84}, 2917, (2000).


\bibitem{Bockrath}
M.~Bockrath, W.~Liang, D.~Bozovic, J.~H. Hafner, C.~M. Lieber, M.~Tinkham, and
  H.~Park.
 {Science} {\bf 291}, 283, (2001).


\bibitem{Kong_transp}
Jing Kong, Erhan Yenilmez, Thomas~W. Tombler, Woong Kim, Hongjie Dai, Robert~B.
  Laughlin, Lei Liu, C.~S. Jayanthi, and S.~Y. Wu.
 {Phys. Rev. Lett.} {\bf 87}, 106801, (2001).


\bibitem{Biel_And_loc}
B.~Biel, F.~J. Garc{\'i}a-Vidal, A.~Rubio, and F.~Flores.
 {Phys. Rev. Lett.} {\bf 95}, 266801, (2005).


\bibitem{Gomez_And_loc}
C.~G{\'o}mez-Navarro, P.~J. de~Pablo, J.~G{\'o}mez-Herrero, B.~Biel, F.~J.
  Garc{\'i}a-Vidal, A.~Rubio, and F.~Flores.
 {Nature Materials} {\bf 4}, 534, (2005).


\bibitem{Son}
Y.-W. Son, J.~Ihm, M.~L. Cohen, S.~G. Louie, and H.~J. Choi.
 {Phys. Rev. Lett.} {\bf 95}, 216602, (2005).


\bibitem{Rocha}
A.~R. Rocha, J.~E. Padilha, A.~Fazzio, and A.~J.~R. da~Silva.
 {Phys. Rev. B} {\bf 77}, 153406, (2008).


\bibitem{Park}
J.-Y. Park.
 {App. Phys. Lett.} {\bf 90}, 023112, (2007).


\bibitem{Bachtold}
Adrian Bachtold, Peter Hadley, Takeshi Nakanishi, and Cees Dekker.
 {Science} {\bf 294}, 1317, (2001).


\bibitem{Carb_elec}
Phaedon Avouris, Zhihong Chen, and Vasili Perebeinos.
 {Nature Nanotechnology} {\bf 2}, 605, (2007).


\bibitem{Mattis}
D.~C. Mattis.
 {Phys. Rev. Lett.} {\bf 32}, 714, (1974).


\bibitem{KaneFisher}
C.~L. Kane and M.~P.~A. Fisher.
 {Phys. Rev. Lett.} {\bf 68}, 1220, (1992).


\bibitem{EggertAffleck}
S.~Eggert and I.~Affleck.
 {Phys. Rev. B} {\bf 46}, 10866, (1992).


\bibitem{kane3}
C.L.~Kane and E.J.~Mele,
Phys. Rev. B {\bf 59}, R12759 (1999).


\bibitem{rommer}
S.~Rommer and S.~Eggert,
Phys. Rev. B {\bf 62}, 4370 (2000).


\bibitem{BalFish}
L.~Balents and M.~P.~A. Fisher.
 {Phys. Rev. B} {\bf 55}, R11973, (1997).


\bibitem{Lin}
H.~H. Lin.
 {Phys. Rev. B} {\bf 58}, 4963, (1998).


\bibitem{relax} A.A.~El-Barbary, R.H.~Telling, C.P.~Ewels, M.I.~Heggi, and P.R.~Birddon, 
Phys.~Rev.~B 68, 144107 (2003).


\bibitem{GogolinNerTsv}
A.~O. Gogolin, A.~A. Nersesyan, and A.~M. Tsvelik.
 {Bosonization and Strongly Correlated Systems}.
 Cambridge University Press, (1998).


\bibitem{White}
S.~R. White.
 {Phys. Rev. Lett.} {\bf 69}, 2863, (1992).


\bibitem{Egger}
R.~Egger and A.~O. Gogolin.
 {Phys. Rev. Lett.} {\bf 79}, 5082, (1997).


\bibitem{Kane2}
C.~Kane, L.~Balents, and M.~P.~A. Fisher.
 {Phys. Rev. Lett.} {\bf 79}, 5086, (1997).


\bibitem{kleiner} 
A.~Kleiner and S.~Eggert, Phys. Rev. B {\bf 64}, 113402 (2001).


\bibitem{meunier}
V.~Meunier and Ph.~Lambin
 {Phys. Rev. Lett.} {\bf 81}, 5588, (1998).


\bibitem{deretz} I.~Deretzis and A.~La Magna, Nanotech. {\bf 17}, 5063 (2006).

\bibitem{loc_state}
States localized at the cap are also possible. Since they have high
  energy in general, we will not consider them here.


\bibitem{SW}
A.~J. Stone and D.~J. Wales.
 {Chem. Phys. Lett.} {\bf 128}, 501, (1986).


\bibitem{strain}
P.~Jensen, J.~Gale, and X.~Blase.
 {Phys. Rev. B} {\bf 66}, 193403, (2002).


\bibitem{Krash_vac}
A.~V. Krasheninnikov, K.~Nordlund, M.~Sirvi{\"o}, E.~Salonen, and J.~Keinonen.
 {Phys. Rev. B} {\bf 63}, 245405, (2001).

\bibitem{graphene}
V.M.~Pereira, J.M.B.~Lopes dos Santos, 
A.H.~Castro Neto,
Phys. Rev. B {\bf 77}, 115109 (2008).



\end{thebibliography}
\end{document}